\documentclass[a4paper,11pt]{article}
\usepackage{amsmath,amssymb,mathrsfs,graphicx,citesort}
\newcommand{\nc}{\newcommand}
\nc{\rnc}{\renewcommand}
\nc{\beq}{\begin{equation}}
\nc{\eeq}{\end{equation}}
\nc{\nn}{\nonumber}
\rnc{\Im}{{\rm{Im}\,}}
\rnc{\Re}{{\rm{Re}\,}}
\rnc{\i}{{\rm i}}
\rnc{\d}{{\rm d}}
\nc{\e}{{\rm e}}
\nc{\mfa}{{\mathfrak{a}}}
\nc{\mfab}{\overline{\mfa}}
\nc{\mfA}{{\mathfrak{A}}}
\nc{\mfAb}{\overline{\mfA}}
\nc{\mfb}{{\mathfrak{b}}}
\nc{\mfbb}{\overline{\mfb}}
\nc{\mfB}{{\mathfrak{B}}}
\nc{\mfBb}{\overline{\mfB}}
\nc{\mfc}{{\mathfrak{c}}}
\nc{\mfcb}{\overline{\mfc}}
\nc{\mfC}{{\mathfrak{C}}}
\nc{\mfCb}{\overline{\mfC}}
\nc{\db}{\displaybreak[0]\\}
\nc{\bra}{\langle}
\nc{\ket}{\rangle}
\nc{\bbra}{\left\langle}
\nc{\kket}{\right\rangle}
\nc{\J}{\mathscr{J}}
\nc{\R}{\mathscr{R}}
\nc{\T}{\mathscr{T}}
\rnc{\H}{\mathscr{H}}
\nc{\M}{\mathscr{M}}
\nc{\Q}{\mathscr{Q}}
\nc{\A}{\mathscr{A}}
\nc{\B}{\mathscr{B}}
\nc{\C}{\mathscr{C}}
\nc{\D}{\mathscr{D}}
\nc{\Z}{\mathcal{Z}}
\nc{\ep}{\varepsilon}
\rnc{\(}{\left(}
\rnc{\)}{\right)}
\nc{\dprime}{\prime\prime}
\nc{\bi}{{\bar{i}}}
\nc{\bT}{{\overline{T}}}
\nc{\bx}{{\widehat{x}}}
\nc{\tx}{{\widetilde{x}}}
\nc{\bm}{{\widehat{m}}}
\nc{\tm}{{\widetilde{m}}}
\nc{\bn}{\widehat{n}}
\nc{\tn}{\widetilde{n}}
\nc{\bp}{\overline{p}}
\nc{\bM}{\overline{M}}
\nc{\bY}{\overline{Y}}
\nc{\bZ}{\overline{Z}}
\nc{\hZ}{\widehat{Z}}
\nc{\tZ}{\widetilde{Z}}
\nc{\tmfb}{\widetilde{\mfb}}
\nc{\bmfb}{\widehat{\mfb}}
\nc{\bGamma}{\widehat{\Gamma}}
\nc{\tGamma}{\widetilde{\Gamma}}

\DeclareMathOperator{\sh}{sh}
\DeclareMathOperator{\ch}{ch}
\DeclareMathOperator{\Tr}{Tr}

\newtheorem{theorem}{Theorem}
\newtheorem{lemma}{Lemma}

\numberwithin{equation}{section}

\begin{document}%
%
\title{Dynamical correlation functions of the 
XXZ model \\ at finite temperature}
\author{
Kazumitsu Sakai\thanks{E-mail:
sakai@gokutan.c.u-tokyo.ac.jp}\\\\
\it Institute of physics, University of Tokyo, \\
\it Komaba 3-8-1,
Meguro-ku, Tokyo 153-8902, Japan \\\\
} 
\date{March 13, 2007}
\maketitle
%
%
\begin{abstract}
Combining a lattice path integral formulation for thermodynamics
with the solution of the quantum inverse scattering problem for
local spin operators, we derive a multiple integral representation 
for the time-dependent longitudinal correlation function of 
the spin-1/2 Heisenberg XXZ chain at finite temperature and 
in an external magnetic field. Our formula reproduces the  
previous results  in the following three limits: the static, 
the zero-temperature and the XY limits.
\end{abstract}

%
%
\section{Introduction}
%
One of the most challenging problems in quantum many-body 
systems, especially in low-dimensions, is related to the  
evaluation of the correlation functions. 
Though many field theoretical schemes or numerical techniques 
have  been developed and achieved remarkable success,
the exact evaluation is in general still very 
difficult even in models which are {\it exactly solvable} 
by the Bethe ansatz methods.

The spin-1/2 Heisenberg XXZ chain is one of the 
simplest but non-trivial solvable models, and has 
served as a testing ground for many theoretical 
approaches. Concerning the correlation function
of this model, Jimbo {\it et al.} \cite{JMMN,JMbook} 
derived a multiple integral representation of 
arbitrary correlators in the off-critical XXZ antiferromagnet
at zero temperature and zero magnetic field. 
Their method utilizing the representation theory of 
the quantum affine algebra $U_q(\widehat{\mathfrak{sl}}_2)$ 
has been extended to the XXX \cite{Naka,KIEU}, 
the massless XXZ \cite{JM}
antiferromagnets. 

Alternative approaches combining the algebraic 
Bethe ansatz and the solution of the quantum inverse 
scattering problem for local spin operators were
proposed by Kitanine {\it et al.} in \cite{KMT00}, 
and were applied to derive  a multiple
integral representation for both the critical and off-critical 
regimes of the XXZ model
in {\it finite} magnetic field.
Moreover they have  obtained  a new multiple integral
representation which is more appropriate for 
the two-point correlators \cite{KMST02,KMST05-1},
and have generalized further to the time-dependent longitudinal
correlation function \cite{KMST05} (see \cite{KMSTreview} for a
recent review).
On the other hand, a finite temperature generalization 
was achieved more recently by G\"ohmann {\it et al.} by
utilizing the quantum transfer matrix formalism for 
thermodynamics \cite{GKS04} (see also \cite{GKS05,GHS05,BGKS} for 
recent progress in this direction).

Motivated by these seminal works, in this paper, we generalize the 
result \cite{KMST05,GKS04} to the time-dependent longitudinal 
correlation function at finite temperature.
By combining a lattice path integral formulation with
the solution of the quantum inverse scattering problem,
we derive a multiple integral representing a generating
function of the time-dependent  correlation 
function for the $z$-components of the spins.
In the zero-temperature, the static and the
XY limits, our formula reproduces the results in
\cite{KMST05,GKS04,CIKT93}.

The layout of this paper is as follows. In the subsequent
section we introduce the spin-1/2 Heisenberg XXZ chain and
its classical counterpart. Utilizing a lattice path integral
formulation together with the solution of the quantum 
inverse scattering problem for local spins, 
in Section~3, we formulate  the generating 
function characterizing the time-dependent longitudinal 
correlation function at finite temperature. In Section~4,
we derive  a multiple integral representing the generating 
function by combining the method developed in \cite{KMST02,KMST05} and
\cite{GKS04}. The following three limits, namely the static,
 the zero-temperature and the XY limits are considered 
in Section~5. Section~6 is devoted to a brief conclusion.

%

%
\section{Spin-1/2 Heisenberg XXZ chain}
%
The Hamiltonian of the spin-1/2 Heisenberg XXZ chain
defined on a periodic lattice of length $L$
is given by
\beq
\H=\H_0-h S^z,
\label{xxz}
\eeq
where
\beq
\H_0=J\sum_{j=1}^L 
  \left\{
  \sigma_j^x\sigma_{j+1}^x+
  \sigma_j^y\sigma_{j+1}^y+
  \Delta(\sigma_j ^z \sigma_{j+1}^z-1)
  \right\}, \quad
S^z=\frac{1}{2}\sum_{j=1}^L \sigma_j^z. 
\eeq
Here $\sigma_j^{x,y,z}$ are the Pauli matrices acting
on the two-dimensional quantum space $\mathcal{H}_j$ at site $j$.  The real 
parameter 
$\Delta_{\ge 0}$ is the anisotropy parameter, and $J \in \mathbb{R}$ fixes the energy 
scale of the model. In this paper, 
we consider the model at finite temperature $T\ge0$ and  
in a finite magnetic filed $h \in \mathbb{R}$.

\begin{figure}[ttt]
\begin{center}
\includegraphics[width=0.85\textwidth]{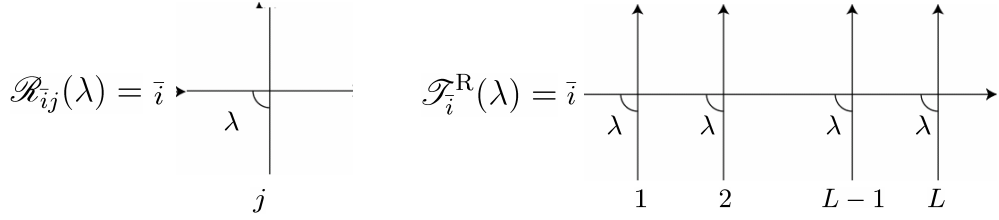}
\end{center}
\caption{A graphical representation for  the $\R$-matrix $\R(\lambda)$
\eqref{r-matrix}
and the monodromy matrix $\T_\bi(\lambda)$ \eqref{monodromy}.}
\label{monodromy-pic}
\end{figure}

It is well-known that a $d$-dimensional quantum system
can be mapped onto  a $(d+1)$-dimensional classical system.
The classical counterpart of the XXZ model is so-called the
six-vertex model whose Boltzmann weights can be described
as the elements of the $\R$-matrix $\R(\lambda)\in
{\rm End}(\mathbb{C}^2\otimes\mathbb{C}^2)$,
\beq
\R(\lambda)=
  \begin{pmatrix}
    1 & 0                & 0                & 0 \\
    0 & \frac{\sh\lambda}{\sh(\lambda+\eta)} 
                   & \frac{\sh \eta}{\sh(\lambda+\eta)}     & 0 \\
    0 & \frac{\sh \eta}{\sh(\lambda+\eta)}  
                   & \frac{\sh \lambda}{\sh(\lambda+\eta)} & 0 \\
    0                & 0                & 0                & 1 
   \end{pmatrix}.
\label{r-matrix}
\eeq
Identifying one of the two vector spaces of the above $\R$-matrix
with the quantum space $\mathcal{H}_j$\footnote{We call the remaining space
the auxiliary space and write $\mathcal{H}_{\bar{i}}$.}, we define the 
monodromy matrix $\T^{\rm R}_\bi(\lambda)$  as
\beq
\T^{\rm R}_{\bi}(\lambda)= \R_{\bi L}(\lambda)\cdots \R_{\bi 2}(\lambda) 
                   \R_{\bi 1}(\lambda),
\label{monodromy}
\eeq
where $\R_{\bi j}(\lambda)$ acts in the space $\mathcal{H}_\bi\otimes\mathcal{H}_j$
(see also figure~\ref{monodromy-pic} for a graphical representation of 
$\R_{\bi j}(\lambda)$ and $\T^{\rm R}_{\bi}(\lambda)$).
Since the $\R$-matrix satisfies the Yang-Baxter equation
\beq
\R_{12}(\lambda-\mu)\R_{13}(\lambda-\nu)\R_{23}(\mu-\nu)=
\R_{23}(\mu-\nu)\R_{13}(\lambda-\nu)\R_{12}(\lambda-\mu),
\label{ybe}
\eeq
the transfer matrix defined by
$
T_{\rm R}(\lambda)=\Tr_{\bi}\T^{\rm R}_{\bi}(\lambda)
$
commutes for different spectral 
parameters: $[T_{\rm R}(\lambda), T_{\rm R}(\mu)]=0$. Using the 
following relation 
connecting the six-vertex model with the XXZ model \eqref{xxz}:
\beq
\H_0=2 J \sh (\eta) T^{-1}_{\rm R}(0)T_{\rm R}'(0), \qquad \Delta= \ch \eta,
\eeq
one can expand the transfer matrix with respect to the spectral parameter
$\lambda$
\beq
T_{\rm R}(\lambda)=
T_{\rm R}(0)\left(1+ \frac{\lambda}{2J\sh \eta}\H_0+\mathcal{O}(\lambda^2)\right).
\label{expandT}
\eeq

For later convenience, let us introduce another type of 
transfer matrix defined by
\beq
\bT_{\rm R}(\lambda)=\Tr_{\bi} \left[\R_{L \bi}(-\lambda)\cdots \R_{2 \bi}(-\lambda)
                       \R_{1 \bi}(-\lambda)\right].
\eeq
Note that $T_{\rm R}(0)$ and $\bT_{\rm R}(0)$ are respectively 
the right- and left-shift operators,
and hence  
\beq
\bT_{\rm R}(0)=T_{\rm R}^{-1}(0).
\label{Tinverse}
\eeq
Using this together with the expansion \eqref{expandT}, 
one arrives at a crucial formula
\beq
\lim_{N\to\infty}\left[
        \bT_{\rm R}(\lambda) T_{\rm R}\(\lambda+\frac{\beta}{N}\)
                 \right]^{\frac{N}{2}} \Biggr|_{\lambda=0}
=\exp\left(\frac{\beta}{4J\sh \eta} \H_0\right), \quad \beta\in \mathbb{C},
\label{trotter}
\eeq
where $N\in 2\mathbb{N}$ is assumed.
%
%
%
%
\section{Time-dependent generating function 
at finite temperature}
%
\subsection{Quantum transfer matrix}
%
Our main aim in this paper is to derive a
multiple integral representation of the 
time-dependent longitudinal
correlation function $\bra \sigma_1^z(0) 
\sigma_{m+1}^z(t)\ket$ for the XXZ model \eqref{xxz}
at finite temperature.
To this end, we firstly consider how 
the time-dependent correlator is described
by the transfer matrix formalism.

\begin{figure}[ttt]
\begin{center}
\includegraphics[width=0.6\textwidth]{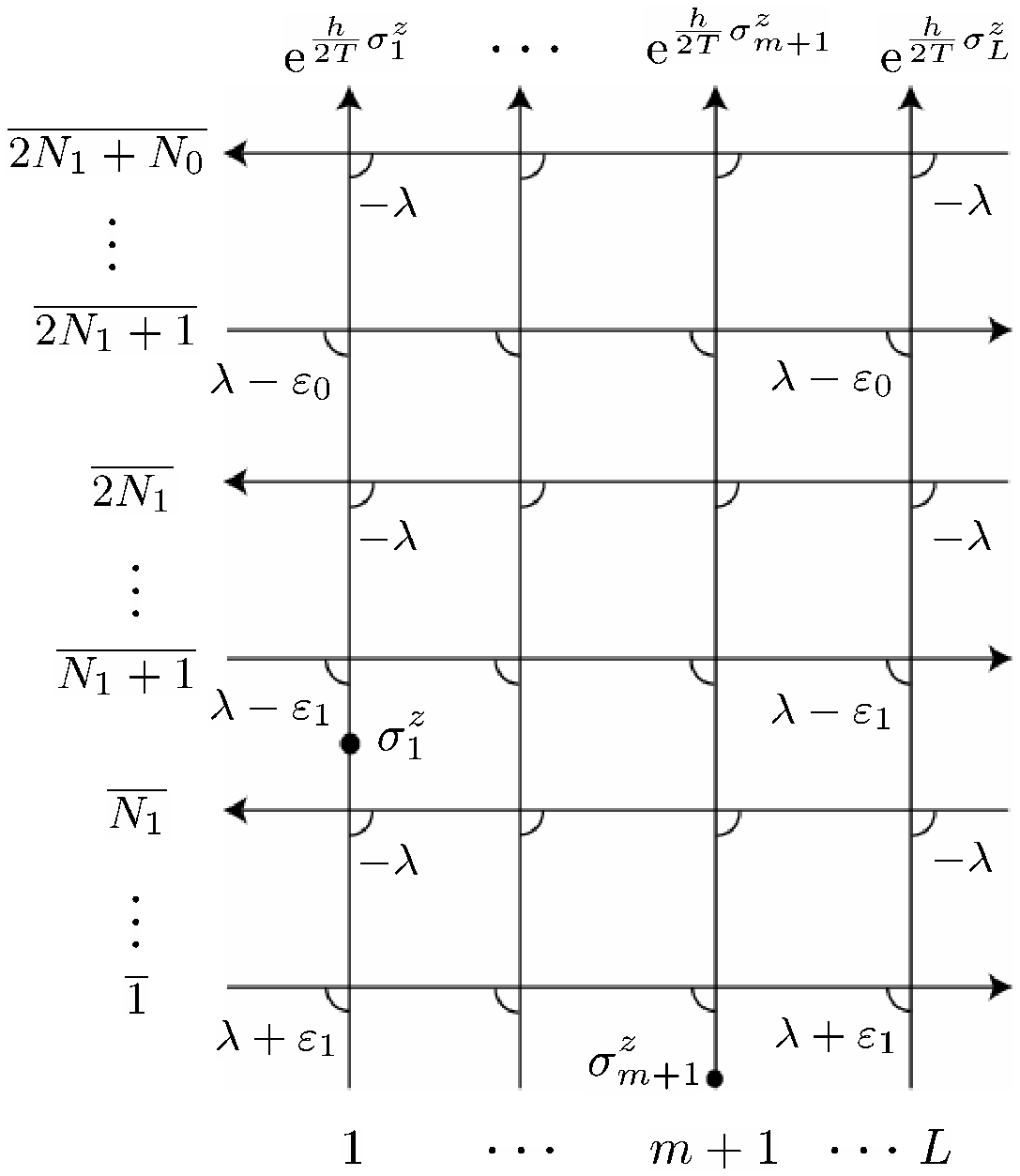}
\end{center}
\caption{A graphical representation for the longitudinal 
dynamical correlation function 
$\bra \sigma_1^z(0)\sigma_{m+1}^z(t) \ket$. 
Here $\ep_0=\beta_0/N_0$ and $\ep_1=
\beta_1/N_1$.
We assume the lattice is on a torus. 
The correlation function (multiplied by the 
partition function $\Tr \e^{-\H/T}$) is reproduced by
setting $\lambda=0$ and taking the limit $N_{0,1}\to\infty$.
The partition function $\Tr \e^{-\H/T}$ is obtained by
just replacing  $\sigma_1^z$ and $\sigma_{m+1}^z$
with the identity operator.
}
\label{dynamical-pic}
\end{figure}

\begin{figure}[ttt]
\begin{center}
\includegraphics[width=0.6\textwidth]{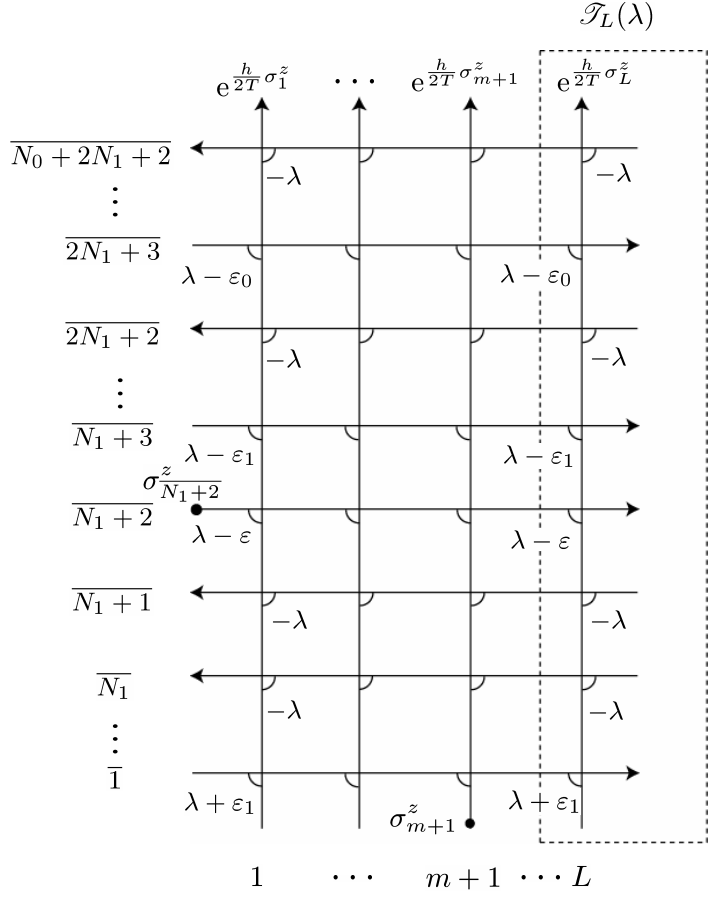}
\end{center}
\caption{A graphical representation of the quantum 
transfer matrix $T(\lambda)=\Tr_j \T_j (\lambda)$ (surrounded by  
broken lines). In fact, a small parameter $\varepsilon$ is introduced
to avoid that the quantum transfer matrix becomes a singular matrix.
The partition function $\Tr \e^{-\H/T}$ is obtained by
replacing  $\sigma_{m+1}^z$ and $\sigma_{\overline{N_1+2}}^z$
with the identity operator.}
\label{qtm-pic}
\end{figure}

Since $[S^z,\sigma_j^z]=0$, the time-dependent local 
spin operator $\sigma_j^z(t)$ can be written as 
\beq
\sigma^{z}_j(t)=\e^{\i \H_0 t}\sigma_j^{z} 
                    \e^{-\i \H_0 t}.
\eeq
Using this and noticing that $[\H_0,S^z]=0$,
one immediately sees
\beq
\bra \sigma_1^z(0)\sigma_{m+1}^z(t)\ket
 =\frac{\Tr \left\{\e^{-\H/T}
            \e^{-\i\H_0 t}\sigma_1^z
            \e^{\i\H_0 t}\sigma_{m+1}^z 
                 \right\}}
         {\Tr \e^{-\H/T}}.
\label{dynamical}
\eeq
Here we have set the Boltzmann constant to unity.
Applying the formula \eqref{trotter}, we insert the relations
\begin{alignat}{2}
& \e^{-\H/T}=\lim_{N_0\to\infty}\e^{\frac{h S^z}{T}} 
      \left\{\bT_{\rm R}(\lambda)T_{\rm R}(\lambda-\ep_0)
      \right\}^{\frac{N_0}{2}} \Bigg|_{\lambda=0}, 
                             & &\qquad  \ep_0=\frac{\beta_0}{N_0},
                                \quad \beta_0=\frac{4 J\sh \eta}{T}, \nn \\
& \e^{\pm \i\H_0 t}=\lim_{N_1\to\infty}
   \left\{\bT_{\rm R}(\lambda)T_{\rm R}(\lambda\pm\ep_1)
      \right\}^{\frac{N_1}{2}} \Bigg|_{\lambda=0},
& &\qquad  \ep_1=\frac{\beta_1}{N_1}, \quad \beta_1=4\i t J  \sh\eta,
\label{Trotter}
\end{alignat}
into \eqref{dynamical}. The result can be graphically 
represented as in figure~\ref{dynamical-pic}.
%
%
%
%
%
As proved in \cite{KMT99,MT00,GK00}, the local spin operator $\sigma_j^z$
is expressed in terms of the transfer matrix:
\beq
\sigma_j^{z}=T_{\rm R}^{j-1}(0) \Tr_{\bi}[ \sigma_{\bi}^{z} 
             \T^{\rm R}_{\bi}(0)] T_{\rm R}^{-j}(0). 
\label{inverse}
\eeq
Thus, substituting
\beq
\sigma^z_1=\lim_{\ep \to 0}\Tr_{\bi}[\sigma_{\bi}^z 
                     \T^{\rm R}_{\bi}(\lambda-\ep)]\bT_{\rm R}(\lambda) 
                     \Bigr|_{\lambda=0},
\quad \ep\in \mathbb{C}
\eeq
which is obtained by \eqref{inverse} and 
\eqref{Tinverse}, one finds 
$\bra \sigma_1^z(0)\sigma_{m+1}^z(t)\ket$ is given by
a product of the transfer matrix and 
two local operators located on the boundaries 
(see figure~\ref{qtm-pic}). 
%
%
%
%

To consider the system at finite temperature,
let us introduce the quantum transfer matrix
$T(\lambda)$ acting in the space 
$\otimes_{\bi=1}^{\overline{N_0+2N_1+2} }
\mathcal{H}_{\bi}$:
\beq
T(\lambda)=\Tr_{j} \T_{j}(\lambda).
\eeq
Here $\T_j(\lambda)$ is defined by
\begin{align}
 \T_j(\lambda)=&
\e^{\frac{h}{2 T}\sigma_j^z}
\R_{j\overline{N_0+2N_1+2}}(-\lambda)
\R_{\overline{N_0+2N_1+1}j}(\lambda-\ep_0)
\cdots
\R_{j\overline{2N_1+4}}(-\lambda)
\R_{\overline{2N_1+3}j}(\lambda-\ep_0) \nn \\
&\times 
\R_{j\overline{2N_1+2}}(-\lambda)
\R_{\overline{2N_1+1}j}(\lambda-\ep_1) 
\cdots
\R_{j\overline{N_1+4}}(-\lambda)
\R_{\overline{N_1+3}j}(\lambda-\ep_1) \nn \\
&\times 
\R_{\overline{N_1+2}j}(\lambda-\varepsilon)
\R_{j\overline{N_1+1}}(-\lambda) \nn \\
&\times 
\R_{j\overline{N_1}}(-\lambda)
\R_{\overline{N_1-1}j}(\lambda+\ep_1) 
\cdots
\R_{j\overline{2}}(-\lambda)
\R_{\overline{1}j}(\lambda+\ep_1).
\end{align}
In figure~\ref{qtm-pic}, we also schematically
depict the quantum transfer matrix.
Thanks to the
Yang-Baxter equation \eqref{ybe}, we see the quantum
transfer matrix  commutes
for different spectral parameters:
$
[T(\lambda),T(\mu)]=0.
$
Thus the dynamical correlation function \eqref{dynamical} is
expressed in terms of $T(\lambda)$:
\beq
\bra\sigma_1^z(0)\sigma_{m+1}^z(t) \ket
=\lim_{N_0,N_1\to\infty}
\lim_{\ep\to0}
\frac{\Tr_{1,\dots, L}\left[T^{L-m-1}(\ep)(A-D)(\ep)
             T^m(\ep) \sigma_{\overline{N_1+2}}^z\right]}
     {\Tr_{1,\dots, L} T^L(\ep)},  
\label{dynamical2}
\eeq
where $A(\lambda)$ and $D(\lambda)$ are  elements
of the monodromy matrix $\T(\lambda)$ represented
as a $2\times 2$ matrix
\beq
\T(\lambda)=
\(  
   \begin{array}{@{\,}cc@{\,}}  
     A(\lambda) & B(\lambda) \\
     C(\lambda) & D(\lambda)
   \end{array}
\)
\eeq
in the quantum space. In this definition, 
$T(\lambda)$
is given by $T(\lambda)=A(\lambda)+D(\lambda)$.

Let us consider the thermodynamic limit $L\to\infty$.
Since the  limits $L\to\infty$ and
$N_0, N_1\to\infty$ are interchangeable as proved in \cite{Suz,SuzIn},
one can take the limit $L\to\infty$ first. 
In addition, we find the leading eigenvalue 
of the quantum transfer matrix $T(0)$ (written as 
$\Lambda_0(0)$) is non-degenerate 
and separated from the next-leading eigenvalues by a 
finite gap even 
in the Trotter limit $N_0, N_1\to\infty$. 
In the thermodynamic limit $L\to\infty$, therefore,
\eqref{dynamical2} is characterized by 
$\Lambda_0(0)$\footnote{Note that $\Lambda_0(\lambda)$ is defined by
$\Lambda_0(\lambda)=\bra\Psi| T(\lambda) |\Psi \ket$.}
and the corresponding (normalized) eigenstate $|\Psi \ket$:
\beq
\bra\sigma_1^z(0)\sigma_{m+1}^z(t) \ket
=\lim_{N_0,N_1\to\infty}\lim_{\ep\to 0}\Lambda^{-m-1}_0(\ep)
\bra \Psi | (A-D)(\ep) T^m(\ep) \sigma_{\overline{N_1+2}}^z
|\Psi \ket.
\eeq
Inserting the relation  
\begin{align}
\sigma_{\overline{N_1+2}}^z
=&
\left[T(-\ep_1) T^{-1}(0)\right]^{\frac{N_1}{2}}
 T^{-1}(0)(A-D)(\ep) 
 T^{-1}(\ep)T(0)
\left[T(0)T^{-1}(-\ep_1) \right]^{\frac{N_1}{2}},
\end{align}
which is a `quantum transfer matrix analogue'
of \eqref{inverse}, we finally obtain
\begin{align}
\bra\sigma_1^z(0)\sigma_{m+1}^z(t) \ket
=\lim_{N_0,N_1\to\infty}\lim_{\ep\to0}
\bra \Psi | & (A-D)(\ep) T^{m-1}(\ep) Q  \nn \\
& \times 
 (A-D)(\ep) Q^{-1} T^{-m-1}(\ep)
 |\Psi \ket,
\label{dynamical-qtm}
\end{align}
where
$
Q=T(\ep)T^{-1}(0)
 \left[T^{-1}(0)T(-\ep_1)\right]^{\frac{N_1}{2}}.
$
%

%
\subsection{Generating function for dynamical correlation function}
%
It is convenient to
introduce the following operator as in \cite{KMST05}:
\beq
\Q^{\kappa}_{l+1, m}=T^{l}(\ep)T_{\kappa}^{m-l}(\ep) Q_{\kappa} T^{-m}(\ep) Q^{-1}.
\label{op-Q}
\eeq
Here $T_{\kappa}(\lambda)$ and
$Q_\kappa$ are respectively defined by 
\beq
T_{\kappa}(\lambda)=A(\lambda)+\kappa D(\lambda),\quad
Q_{\kappa}=T_{\kappa}(\ep)T^{-1}_{\kappa}(0)
 \left[T^{-1}_{\kappa}(0)T_{\kappa}(-\ep_1)\right]^{\frac{N_1}{2}}.
\eeq
Due to the Yang-Baxter equation \eqref{ybe}, 
the twisted quantum transfer matrix 
$T_{\kappa}(\lambda)$ is commutative as 
long as the twist angle $\kappa$ is taken 
the same:
$
[T_{\kappa}(\lambda),T_{\kappa}(\mu)]=0.
$
It follows that
\begin{align}
\bbra \frac{1-\sigma_{l+1}^z(0)}{2}
          \frac{1-\sigma_{m+1}^z(t)}{2}
\kket 
&=\lim_{N_0,N_1\to\infty}\lim_{\ep\to0}
 \bra \Psi|T^{l}(\ep) D(\ep) T^{m-l-1}(\ep) Q \nn \\
&\qquad\qquad\qquad\qquad\qquad \times D(\ep)Q^{-1}T^{-m-1}(\ep) |\Psi \ket \nn \\
&\hspace*{-3.2cm}=\lim_{N_0,N_1\to\infty}\lim_{\ep\to0}\frac{1}{2}
 \partial_{\kappa}^2\bra \Psi|\Q^{\kappa}_{l+1,m+1}-
                    \Q_{l+2,m+1}^{\kappa}
                   -\Q_{l+1,m}^{\kappa}+\Q^{\kappa}_{l+2,m}
|\Psi \ket \bigl|_{\kappa=1}.
\end{align}
Because of the translational invariance for the
correlation functions, one can set $l=0$ without
loss of generality. Introducing the  
generating function
\beq
\mathcal{Q}_{\kappa}(m,t)=\lim_{N_0,N_1\to\infty}\lim_{\ep\to0}
\bra \Psi|\Q^{\kappa}_{1,m}|\Psi \ket,
\label{g-function}
\eeq
one obtains the longitudinal time-dependent correlation 
function \eqref{dynamical}:
\beq
\bra \sigma_1^z(0)\sigma_{m+1}^z(t) \ket
=2 D_{ m}^2 \partial_\kappa^2 \mathcal{Q}_\kappa(m,t)\bigl|_{\kappa=1}
  +2\bra\sigma^z \ket-1,
\label{op-Q2}
\eeq
where $\bra \sigma^z \ket$ is the magnetization (multiplied by a factor 2) 
and $D_m$ denotes the lattice derivative defined
by 
\beq
D_m g(m)=g(m+1)-g(m), \quad D^2_mg(m)= g(m+1)-2g(m)+g(m-1),
\quad m\in \mathbb{N}.
\eeq
%
\subsection{Bethe ansatz}
%
To evaluate \eqref{g-function} actually, we need
to investigate the leading eigenvalue and the corresponding
eigenstate. Here we derive a general formula describing the 
eigenvalues and the corresponding eigenstates 
through the solutions to a certain algebraic 
equation called the Bethe ansatz equation.

Let us define the `vacuum state'
$|0\ket$ as
\begin{align}
|0\ket:=&\underbrace{\binom{1}{0}_{\overline{1}}
          \otimes
        \binom{0}{1}_{\overline{2}}
          \otimes \cdots
          \otimes
        \binom{0}{1}_{\overline{N_1}}}_{N_1 {\rm \, factors}}
          \otimes
        \binom{0}{1}_{\overline{N_1+1}}
          \otimes
        \binom{1}{0}_{\overline{N_1+2}} \nn \\
&       \underbrace{\otimes \binom{1}{0}_{\overline{N_1+3}} 
          \otimes
        \binom{0}{1}_{\overline{N_1+4}}
        \otimes \cdots
        \otimes
        \binom{0}{1}_{\overline{2N_1+2}}}_{N_1 {\rm \, factors}} \nn \\
&   \underbrace{\otimes \binom{1}{0}_{\overline{2N_1+3}} 
          \otimes
        \binom{0}{1}_{\overline{2N_1+4}}
        \otimes \cdots
        \otimes
        \binom{0}{1}_{\overline{N_0+2N_1+2}} }_{N_0 {\rm \, factors}}.
\label{vacuum}
\end{align}
Obviously \eqref{vacuum} is an eigenstate of the (twisted) 
quantum transfer matrix $T_{\kappa}(\lambda)$. Explicitly
\beq
T_{\kappa}(\lambda)|0\ket=(a(\lambda)+\kappa d(\lambda))|0\ket,
\quad A(\lambda)|0\ket=a(\lambda)|0\ket,
\quad D(\lambda)|0\ket=d(\lambda)|0\ket,
\eeq
where
\begin{align}
a(\lambda)&=\left\{\frac{\sh \lambda}
                 {\sh(\lambda-\eta)}\right\}^{\frac{N_0}{2}+N_1+1}
\e^{\frac{h}{2T}}, \nn \\
d(\lambda)& =\left\{\frac{\sh(\lambda+\ep_1)}
              {\sh(\lambda+\ep_1+\eta)}
                  \frac{\sh(\lambda-\ep_1)}
              {\sh(\lambda-\ep_1+\eta)}\right\}^{\frac{N_1}{2}}
           \left\{\frac{\sh(\lambda-\ep_0)}
              {\sh(\lambda-\ep_0+\eta)}\right\}^{\frac{N_0}{2}}
            \frac{\sh(\lambda-\ep)}
                 {\sh(\lambda-\ep+\eta)}
      \e^{-\frac{h}{2T}}.
\end{align}
In the framework of the algebraic Bethe ansatz, the
vector $|\{\lambda\}_{\kappa}\ket$ 
constructed by the multiple action of $B(\lambda)$, namely
$
|\{\lambda^\kappa\}\ket=\prod_{j=1}^M B(\lambda^\kappa_j)|0\ket
$,
is an eigenstate of $T_\kappa(\lambda)$ if the rapidities 
$\{\lambda_j^\kappa\}_{j=1}^M$
satisfy the following Bethe ansatz equation:
\beq
\frac{a(\lambda_j^\kappa)}{d(\lambda_j^\kappa)}=-\kappa\prod_{k=1}^M 
                      \frac{\sh(\lambda^\kappa_j-\lambda^\kappa_k+\eta)}
                           {\sh(\lambda^\kappa_j-\lambda^\kappa_k-\eta)}.
\label{bae}
\eeq
The corresponding eigenvalue is given by
\beq
\Lambda(\lambda,\kappa)=
a(\lambda)\prod_{j=1}^M\frac{\sh(\lambda-\lambda_j^\kappa-\eta)}
                                                {\sh(\lambda-\lambda_j^\kappa)}+
                    \kappa d(\lambda)\prod_{j=1}^M\frac{\sh(\lambda-\lambda_j^\kappa+\eta)}
                                                {\sh(\lambda-\lambda_j^\kappa)}.
\label{eigenvalue}
\eeq

\begin{figure}[ttt]
\begin{center}
\includegraphics[width=0.49\textwidth]{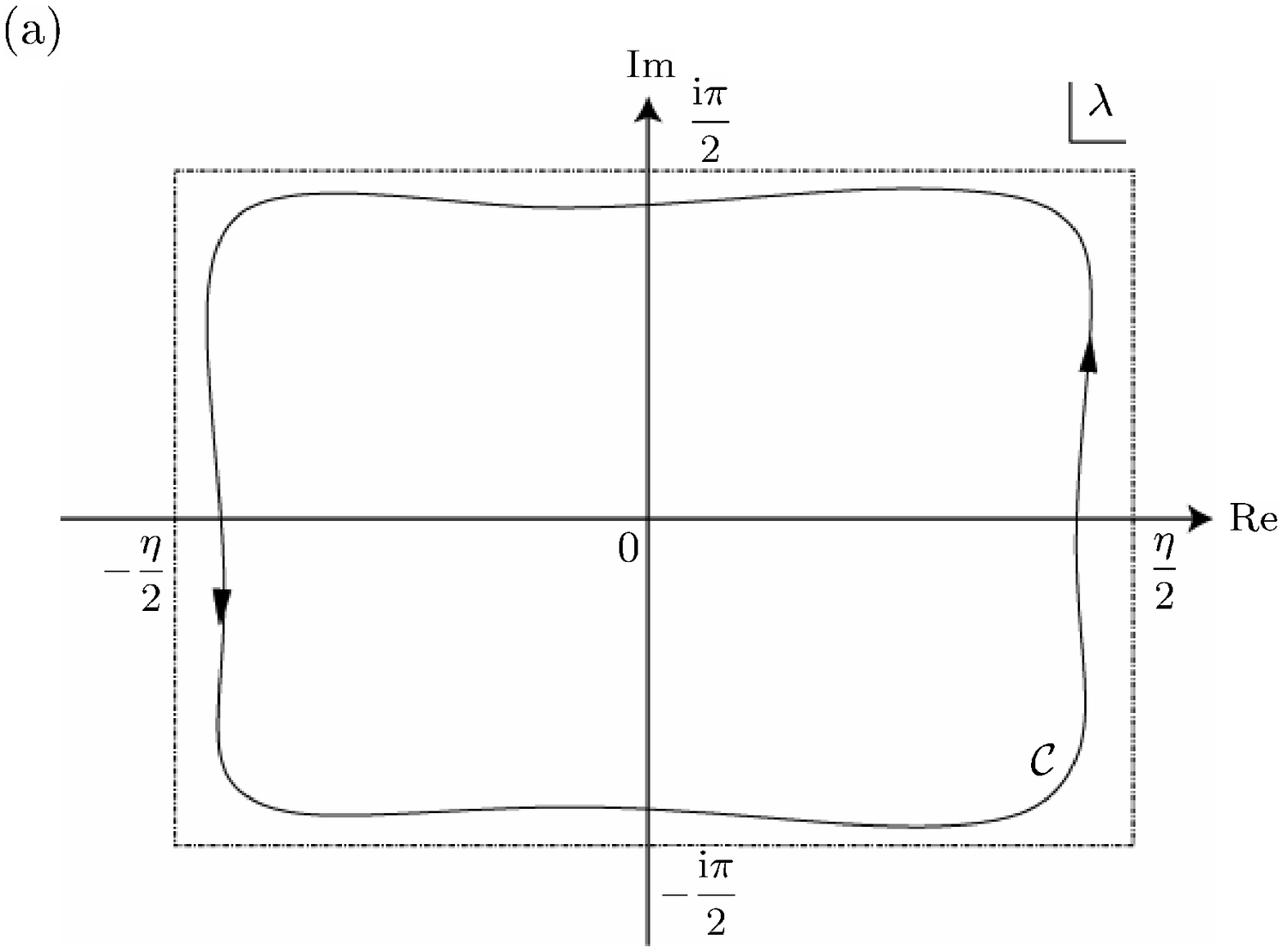}
\includegraphics[width=0.49\textwidth]{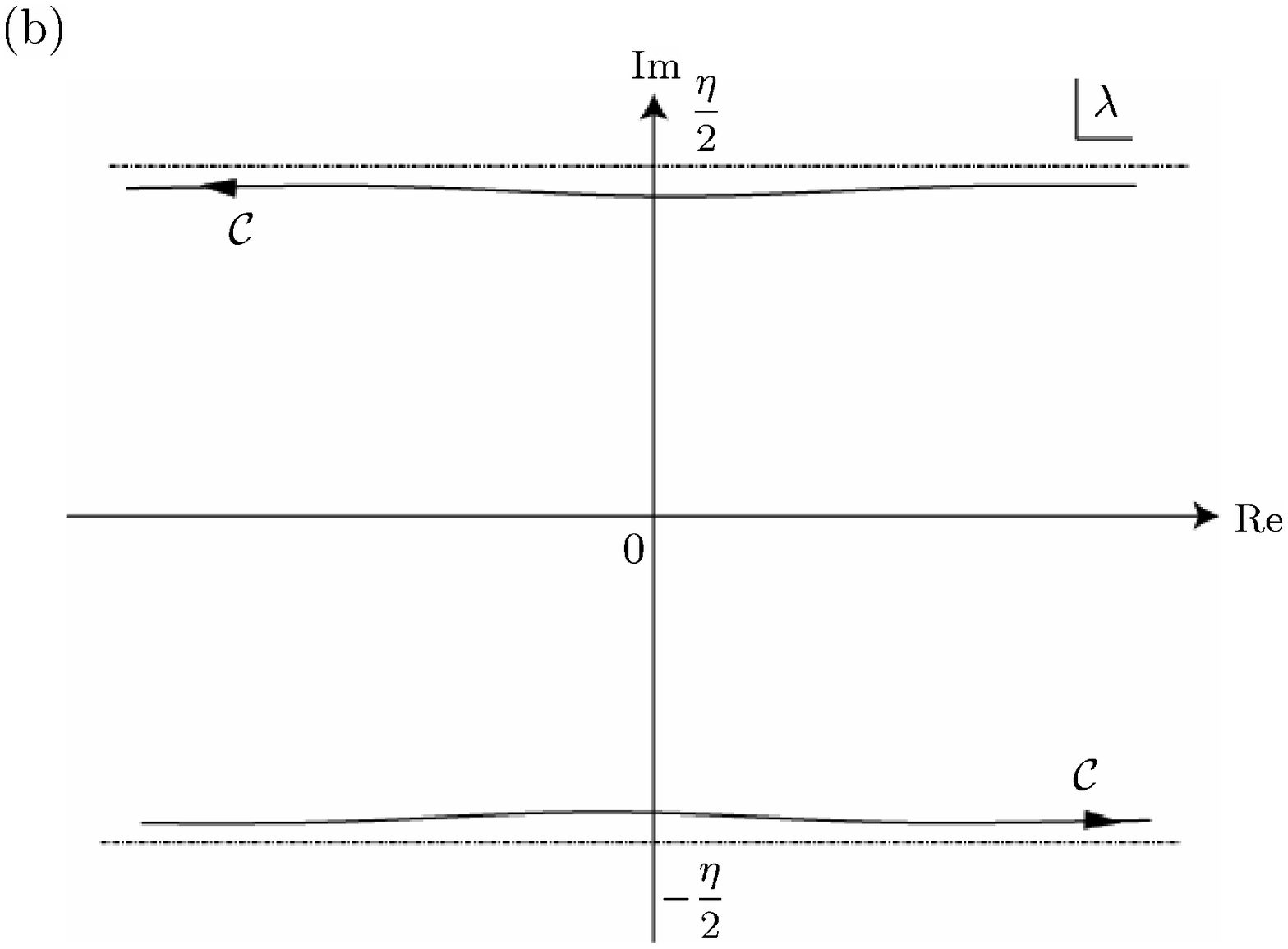}
\end{center}
\caption{The integration contours for the off-critical
regime $\Delta>1$ (a) and for the critical regime
$0\le \Delta \le 1$ (b). }
\label{contour-pic}
\end{figure}

Hereafter we restrict ourselves on the case $\kappa=1$ in
\eqref{bae} and simply write the roots and the eigenstate
corresponding to the leading eigenvalue $\Lambda_0(0)(=\Lambda_0(0,1))$ as
$\{\lambda_j\}_{j=1}^{N/2}$ ($N=N_0+2N_1+2$)
and $|\{\lambda\}\ket$, respectively.
To make the analysis possible even in the
Trotter limit $N\to\infty$, we utilize a powerful
method  as in \cite{Klumper92,Klumper93,Destri92}.
Let us consider the following auxiliary function
\beq
\mfa(\lambda)=
\frac{d(\lambda)}{a(\lambda)}\prod_{k=1}^{\frac{N}{2}} 
                      \frac{\sh(\lambda-\lambda_k+\eta)}
                           {\sh(\lambda-\lambda_k-\eta)},
\quad N=N_0+2N_1+2,
\eeq
which associates the Bethe ansatz roots $\{\lambda_j\}_{j=1}^{N/2}$
with zeros of $1+\mfa(\lambda)$. By studying the analyticity 
properties of the auxiliary function, one sees $\mfa(\lambda)$
satisfies the following nonlinear
integral equation:
\begin{align}
\ln\mfa(\lambda)=&-\frac{h}{T}+\ln\frac{\sh(\lambda+\eta)\sh(\lambda-\ep)}
        {\sh(\lambda)\sh(\lambda-\ep+\eta)}+
\frac{N_0}{2}
\ln\frac{\sh(\lambda+\eta)\sh(\lambda-\ep_0)}
        {\sh(\lambda)\sh(\lambda-\ep_0+\eta)}
        \nn \\
&+\frac{N_1}{2}
\ln\frac{\sh(\lambda+\eta)\sh(\lambda+\ep_1)} 
        {\sh(\lambda)\sh(\lambda+\ep_1+\eta)}
+\frac{N_1}{2}
\ln\frac{\sh(\lambda+\eta)\sh(\lambda-\ep_1)} 
        {\sh(\lambda)\sh(\lambda-\ep_1+\eta)} \nn \\
& 
-\int_{\mathcal{C}}\frac{\d\omega}{2\pi\i}\frac{\sh(2\eta)\ln(1+\mfa(\omega))}
             {\sh(\lambda-\omega+\eta)\sh(\lambda-\omega-\eta)}. 
\label{nlie-f}
\end{align} 
Here the contour $\mathcal{C}$ is taken, for instance, as a
rectangular contour whose edges are parallel to the
real axis at $\pm \pi \i/2$ 
(respectively $\pm \eta/2$) and are parallel
to the imaginary axis at $\pm \eta/2$ (respectively
$\pm \infty$) for
the off-critical regime $\Delta=\ch\eta>1$ (respectively for
the critical regime $0\le \Delta=\ch\eta\le 1$)
(see figure~\ref{contour-pic} for a pictorial definition).
In \eqref{nlie-f}
the limits $N_0,N_1\to\infty$ and $\ep\to 0$
can be taken analytically. 
We thus obtain 
\beq
\ln\mfa(\lambda)=-\frac{h}{T}-\frac{2 J \sh^2(\eta)}
{T\sh(\lambda)\sh(\lambda+\eta)}-
\int_{\mathcal{C}}\frac{\d\omega}{2\pi\i}\frac{\sh(2\eta) 
                      \ln(1+\mfa(\omega))}
             {\sh(\lambda-\omega+\eta)\sh(\lambda-\omega-\eta)},
\label{nlie1}
\eeq
which is exactly the same as that in \cite{GKS04}. For later convenience,
here we also introduce another auxiliary function $\mfab(\lambda)
=1/\mfa(\lambda)$ satisfying the following 
nonlinear integral equation in the limits $N_0,N_1 \to\infty$
and $\ep\to 0$  \cite{GKS04}:
\beq
\ln\mfab(\lambda)=\frac{h}{T}-\frac{2 J \sh^2(\eta)}
{T\sh(\lambda)\sh(\lambda-\eta)}+
\int_{\mathcal{C}}\frac{\d\omega}{2\pi\i}\frac{\sh(2\eta) 
                      \ln(1+\mfab(\omega))}
             {\sh(\lambda-\omega+\eta)\sh(\lambda-\omega-\eta)}.
\label{nlie2}
\eeq

By this auxiliary
function $\mfa(\lambda)$, the leading eigenvalue of the
quantum transfer matrix related to the free energy density $f$
by $f=-T\ln\Lambda_0(0)$ is expressed as a single integral
form
\beq
\ln\Lambda_0(0)=\frac{h}{2T}+\int_{\mathcal{C}}
      \frac{\d\omega}{2\pi\i}
        \frac{\sh(\eta)\ln(1+\mfa(\omega))}
             {\sh(\omega)\sh(\omega+\eta)}
=-\frac{h}{2T}-\int_{\mathcal{C}}
      \frac{\d\omega}{2\pi\i}
        \frac{\sh(\eta)\ln(1+\mfab(\omega))}
             {\sh(\omega)\sh(\omega-\eta)}.
\label{largest}
\eeq
Differentiating \eqref{largest} with respect to $h$, 
one has the magnetization (multiplied by a factor 2),
\begin{equation}
\bra \sigma^z \ket=1+T\int_{\mathcal{C}}
      \frac{\d\omega}{\pi\i}
        \frac{\sh(\eta)  \partial_h\mfa(\omega)}
             {\sh(\omega)\sh(\omega+\eta)(1+\mfa(\omega))}
 =-1-T\int_{\mathcal{C}}
      \frac{\d\omega}{\pi\i}
        \frac{\sh(\eta) \partial_h\mfab(\omega)}
             {\sh(\omega)\sh(\omega-\eta)(1+\mfab(\omega))}.
\label{magnet}
\end{equation}
%
%
%
%
%
\section{Multiple integral representation}
%
%
In this section, using the method developed in
\cite{GKS04}, we derive a multiple integral
representation of the longitudinal dynamical
correlation function 
$\bra \sigma_1^z(0)\sigma_{m+1}^z(t)\ket$ at
finite temperature.
Utilizing the relation 
$
T_{\kappa}(0)T_{\kappa}(\eta)=\kappa a(\eta)d(0),
$
which can easily be obtained from \eqref{eigenvalue},
one expresses the operator $\Q_{1,m}^{\kappa}$ \eqref{op-Q}
as 
\begin{align}
\Q_{1,m}^\kappa=\kappa^{-\frac{N_1}{2}-1}&
[T_{\kappa}^{m+1}(\ep)T_{\kappa}^{\frac{N_1}{2}}(-\ep_1)]
T_{\kappa}^{\frac{N_1}{2}+1}(\eta)
[T^{-m-1}(\ep)T^{-\frac{N_1}{2}}
(-\ep_1)]T^{-\frac{N_1}{2}-1}(\eta).
\end{align}
To analyze the above quantity, we conveniently
introduce a set of  parameters
$\xi_1,\dots,\xi_{\bm+\tm}$ ($\bm=
N_1/2+m+1$, $\tm=N_1/2+1$) located
inside $\mathcal{C}$ and define 
\begin{align}
(x_1,\dots,  x_{\bm+\tm})
&=(\bx_1,\dots,\bx_{\bm};
  \tx_1,\dots,\tx_{\tm}) \nn \\
&=(\xi_1+\ep,\dots,
   \xi_{m+1}+\ep,
   \xi_{m+2}-\ep_1,\dots,
   \xi_{\bm}-\ep_1;
   \xi_{\bm+1}+\eta,\dots,
   \xi_{\bm+\tm}+\eta).
\label{inhom}
\end{align}
It follows that
\begin{align}
&\bra\Psi| \Q_{1,m}^\kappa |\Psi \ket
=\lim_{\xi_1,\dots,\xi_{\bm+\tm}\to 0}
  \Phi_N(\kappa|\{\xi\}) \nn \\
&\Phi_N(\kappa|\{\xi\})=
\kappa^{-\tm}\frac{\bra \{\lambda\}|
 \prod_{j=1}^{\bm+\tm}
         T_{\kappa}(x_j)
 \prod_{j=1}^{\bm+\tm}
         T^{-1}(x_j)
 |\{\lambda\} \ket}
{\bra \{\lambda\}|\{\lambda\} \ket},
\label{Phi}
\end{align}
where $\{\lambda\}$ and $|\{\lambda\}\ket$
are, respectively, the Bethe ansatz roots and 
the eigenstate corresponding to the 
leading eigenvalue $\Lambda_0(0)$
(see the previous section). Note that 
the dual vector $\bra \{\lambda\}|$ is 
constructed by  the multiple action of 
$C(\lambda)$ on the  state $\bra 0|$ which
is the transposition of the vacuum state $|0\ket$. Namely
$
\bra \{\lambda\}|=\bra 0| \prod_{j=1}^{N/2} C(\lambda_j)
$, 
where $N=N_0+2N_1+2$.

Expression \eqref{Phi} is  formally
similar to equation (75) in \cite{GKS04}.
The essential difference only lies in
the definition of the parameters $\{x\}$ \eqref{inhom}.
In the present case, 
(i) the number of parameters $\{ x \}$ 
and elements of  $\{\bx\}$
explicitly depend on the Trotter number $N_1$;
(ii) in the homogeneous limit
$\xi_j\to 0$, the elements of 
$\{\tx\}$ converge to $\eta$ 
which is outside the contour $\mathcal{C}$.
On the other hand, for the static correlation function
\cite{GKS04}, the number of the parameters is 
equal to the distance of the correlator i.e. $m$. 
Furthermore all the parameters converges to  0
(i.e. inside $\mathcal{C}$) in the 
homogeneous limit $\xi_j\to0$.

To extend the formulation in \cite{GKS04}
to the time-dependent case, we first introduce
the following  lemma, which is still applicable
to the present case.
\begin{lemma} \label{GKS-lemma} \cite{GKS04}. $\Phi_N(\kappa|\{\xi\})$
\eqref{Phi}
has the following representation as a sum over partitions
of the Bethe ansatz roots $\{\lambda\}$, and of the
inhomogeneous parameters $\{x\}$ \eqref{inhom} (or equivalently $\{\xi\}$).
\beq
\Phi_N(\kappa|\{\xi\})=\kappa^{-\tm}\sum_{n=0}^{\bm+\tm}
           \sum_{\substack{\{\lambda \}=\{\lambda^+ \} \cup \{\lambda^-\} \\
                 \{ x \}=\{x^+ \}\cup \{x^- \} \\ 
                 |\lambda^+|=|x^+|=n}}
           \frac{\bY_{|x^+|}(\{\lambda^+\}|\{x^+\})
                 \bZ_{|x^+|}(\{\lambda^+\}|\{x\})}
                {\prod_{j=1}^{|x^+|}\mfa'(\lambda_j^+)
                 \prod_{j=1}^{|x^-|}(1+\mfa(x_j^-))},
\label{Phi-lemma}
\eeq
where $|\lambda|$, $|x|$, etc denote the number of elements of 
$\{\lambda\}$, $\{x\}$, etc. The  two functions 
$\bY_n(\{\lambda^+\}|\{x^+\})$ and $\bZ_n(\{\lambda^+\}|\{x\})$ are
respectively defined by 
\begin{align}
&\bY_n(\{\lambda^+\}|\{x^+\})=\prod_{j=1}^n \left[
   \frac{\mfb_+(\lambda_j^+)}{\mfb_+'(x_j^+)}
   \prod_{k=1}^n \frac{\sh(\lambda_j^+ - x_k^+ +\eta)
                         \sh(\lambda_j^+ - x_k^+ -\eta)}
                        {\sh(x_j^+ - x_k^+ +\eta)
                         \sh(\lambda_j^+ - \lambda_k^+-\eta)} \right] \nn \\
&\qquad \qquad \qquad \qquad \times
 \det \bM(\lambda_j^+,x_k^+)\det G(\lambda_j^+,x_k^+), \nn \\
&\bZ_n(\{\lambda^+\}|\{x\})=\prod_{j=1}^{\bm+\tm-n}\left[
  1+\kappa \mfa(x_j^-)\prod_{k=1}^n \frac{f(x_j^-,x_k^+)f(\lambda_k^+,x_j^-)}
                                         {f(x_k^+,x_j^-)f(x_j^-,\lambda_k^+)}
  \right],
\end{align}
where
\begin{align}
&\mfb_{\pm}(\lambda)=\prod_{k=1}^{\bm+\tm} \frac{\sh(\lambda-x_k)}
                                          {\sh(\lambda-x_k\pm\eta)},
\quad
f(\lambda,\mu)=\frac{\sh(\lambda-\mu+\eta)}{\sh(\lambda-\mu)}, \nn \\
&\bM(\lambda^+_j,x^+_k)=t(x_k^+,\lambda_j^+)+\kappa t(\lambda_j^+,x_k^+)
            \prod_{l=1}^n\frac{\sh(\lambda_j^+ -\lambda_l^+-\eta)
                               \sh(\lambda_j^+ -x_l^++\eta)}
                              {\sh(\lambda_j^+ -\lambda_l^+ +\eta)
                               \sh(\lambda_j^+ -x_l^+-\eta)},
\label{M-function}
\end{align}
and $G(\lambda,x)$ is the solution of a
linear integral equation
\begin{align}
G(\lambda,x)=&t(x,\lambda)+\int_{\mathcal{C}} \frac{\d \omega}{2\pi \i}
        \frac{\sh(2\eta)}{\sh(\lambda-\omega+\eta)\sh(\lambda-\omega-\eta)}
  \frac{G(\omega,x)}{1+\mfa(\omega)}   \nn \\
&-
\begin{cases}
0 & \text{for $x\in\{\bx\}$} \\
\left\{\frac{\sh(2\eta)}{\sh(\lambda-x+\eta)
                               \sh(\lambda-x-\eta)}+
              \frac{\sh(2\eta)}{\sh(\lambda-x)\sh(\lambda-x+2\eta)}
               \frac{1}{1+\mfa(x-\eta)}\right\}\frac{1}{1+\mfa(x)}
 & \text{for $x\in\{\tx\}$}
\label{G-function}
\end{cases}
\end{align}
with
\beq
t(\lambda,\mu)=\frac{\sh(\eta)}{\sh(\lambda-\mu)\sh(\lambda-\mu+\eta)}.
\eeq
\end{lemma}
Note that the range of the variable $x$ in 
\eqref{G-function} is extended from inside the contour 
$\mathcal{C}$ to outside $\mathcal{C}$ by analytic 
continuation.

In order to proceed to the dynamical case, let us
modify \eqref{Phi-lemma}
in Lemma~\ref{GKS-lemma}. After  simple calculations, we obtain
\begin{align}
\Phi_N(\kappa|&\{\xi\})=\sum_{\bn=0}^{\bm}
           \sum_{\tn=0}^{\tm} 
           \sum_{\substack{\{\lambda \}=\{\lambda^+ \} \cup \{\lambda^-\} \\
                 |\lambda^+|=n=\bn+\tn}}
            \sum_{\substack{
                 \{ \bx \}=\{\bx^+ \}\cup \{\bx^- \} \\ 
                 |\bx^+|=\bn}} 
           \sum_{\substack{
                 \{ \tx \}=\{\tx^+ \}\cup \{\tx^- \} \\ 
                 |\tx^+|=\tn}}                      \nn \\
& \left[\kappa^{-\tn} \frac{Y_{|\bx^+|,|\tx^+|}(\{\lambda^+\}|\{x^+\})
                 \hZ_{|\bx^+|,|\tx^+|}(\{\lambda^+\}|\{x\})
                 \tZ_{|\bx^+|,|\tx^+|}(\{\lambda^+\}|\{x\})}
                {\prod_{j=1}^{|x^+|}\mfa'(\lambda_j^+)
                 \prod_{j=1}^{|\bx^-|}(1+\mfa(\bx_j^-))
                 \prod_{j=1}^{|\tx^-|}(1+\mfab(\tx_j^-))}
\right],
\label{Phi-mod}
\end{align}
where $\bn$ and $\tn$, respectively, denote the number 
of  elements of $\{\bx^+\}$ and $\{\tx^+\}$ (i.e.
$\bn=|\bx^+|$; $\tn=|\tx^+|$), $n=\bn+\tn$, and
\begin{align}
Y_{\bn,\tn}(\{\lambda^+\}|\{x^+\})=&
\prod_{j=1}^{n} \left[
   \frac{\bmfb_+(\lambda_j^+)\tmfb_-(\lambda_j^+)}
        {\bmfb_+'(x_j^+)\tmfb_-'(x_j^+)} 
      \prod_{k=1}^{\bn} \frac{\sh(\lambda_j^+ - \bx_k^+ +\eta)}
                        {\sh(x_j^+ - \bx_k^+ +\eta)} 
      \prod_{k=1}^{\tn} \frac{\sh(\lambda_j^+ - \tx_k^+ -\eta)}
                        {\sh(x_j^+ - \tx_k^+ -\eta)} \right] \nn \\
    & \times \prod_{j,k=1}^n \left[
      \frac{\sh(\lambda_j^+ -x_k^+-\eta)}{\sh(\lambda_j^+-\lambda_k^+-\eta)}\right]
       \det \bM(\lambda_j,x_k)\det G(\lambda_j,x_k), 
                                                        \nn \\
\hZ_{\bn,\tn}(\{\lambda^+\}|\{x\})=&\prod_{j=1}^{\bm-\bn}
     \left[
     1+\kappa \mfa(\bx_j^-)\prod_{k=1}^n 
                  \frac{f(\bx_j^-,x_k^+)f(\lambda_k^+,\bx_j^-)}
                       {f(x_k^+,\bx_j^-)f(\bx_j^-,\lambda_k^+)}                      
     \right], \nn \\
\tZ_{\bn,\tn}(\{\lambda^+\}|\{x\})=&\prod_{j=1}^{\tm-\tn}
\left[
  1+\kappa^{-1} \mfab(\tx_j^-)\prod_{k=1}^n \frac{f(x_k^+,\tx_j^-)f(\tx_j^-,\lambda_k^+)}
                                           {f(\tx_j^-,x_k^+)f(\lambda_k^+,\tx_j^-)}
  \right].
\end{align}
Here $\bmfb_{\pm}(\lambda)$ and  $\tmfb_{\pm}(\lambda)$ denote
\beq
\bmfb_\pm(\lambda)=\prod_{k=1}^{\bm}\frac{\sh(\lambda-\bx_k)}{\sh(\lambda-\bx_k\pm\eta)},\quad
\tmfb_\pm(\lambda)=\prod_{k=1}^{\tm}\frac{\sh(\lambda-\tx_k)}{\sh(\lambda-\tx_k\pm\eta)}.
\eeq

\begin{figure}[ttt]
\begin{center}
\includegraphics[width=0.6\textwidth]{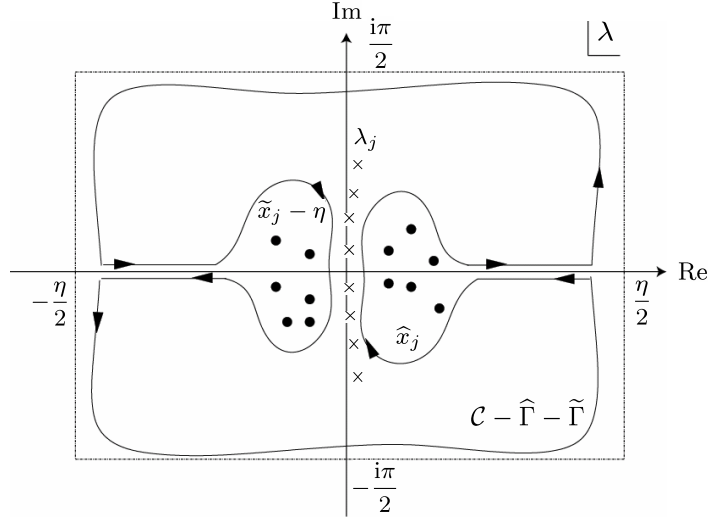}
\end{center}
\caption{The integration contour $\mathcal{C}-\bGamma-\tGamma$ corresponding
to the off-critical regime $\Delta>1$.}
\label{contour-pic2}
\end{figure}

The remaining task is to replace the sums over
the partitions of the set $\{\lambda\}$ and $\{x\}$
in \eqref{Phi-mod} with a certain set of contour 
integrals, where the Trotter limit
 will be taken analytically. Consider first the
partitions for the set of the Bethe ansatz roots 
$\{\lambda\}$.
Let $f(\omega_1,\dots,\omega_n)$ be  analytic on and inside the contour $\mathcal{C}$,
symmetric with respect to $n$ variables $\omega_j$, and zero when 
any two of its variables are the same.
The poles  of
the function $1/(1+\mfa(\omega))$ inside 
$\mathcal{C}$ are simple poles at 
$\omega=\lambda_j$ with
residues $1/\mfa'(\lambda_j)$ ($j=1,\dots,N/2$).
Hence the following is valid.
\beq
\frac{1}{n!}\int_{\mathcal{C}^n}\prod_{j=1}^n
       \frac{\d\omega_j}{2\pi\i(1+\mfa(\omega_j))}
       f(\omega_1,\dots,\omega_n)=
   \sum_{\substack{\{\lambda\}=\{\lambda^+\}\cup\{\lambda^-\} \\
                   |\lambda^+|=n}} 
    \frac{f(\lambda_1^+,\dots,\lambda_n^+)}{\prod_{j=1}^n \mfa'(\lambda_j^+)}.
\label{sum2int}
\eeq
Note here that the relation similar to the above is also
holds for $\mfab(\omega)$. If the summand in \eqref{Phi-mod}
is considered to be a function of $n$ variables 
$\{\lambda_j^+\}_{j=1}^{n}$, one finds it has simple 
poles at $\lambda_j^+=\bx_k^+$ and 
$\lambda_j^+=\tx_k^{\pm}-\eta$.
Since the parameters $\{\xi\}$ \eqref{inhom} can
be chosen arbitrary values inside $\mathcal{C}$, 
we choose $\{x\}$ such that the two sets of 
parameters $\{x\}$ and $\{\lambda\}$ are distinguishable.
Then there exists a simple closed contour 
surrounding the Bethe ansatz roots $\{\lambda\}$
but excluding $\{x\}$ (see figure~\ref{contour-pic2}).
Let $\mathcal{C}-\bGamma-\tGamma$ be such a contour,
where $\bGamma$ (respectively $\tGamma$) encircles $\{\bx \}$
(respectively $\{\tx \}$).
Applying \eqref{sum2int} into \eqref{Phi-mod}, we obtain
\begin{align}
&\sum_{\substack{\{\lambda \}=\{\lambda^+ \} 
                    \cup \{\lambda^-\} \\
                 |\lambda^+|=n}}
 \frac{Y_{\bn,\tn}(\{\lambda^+\}|\{x^+\})
                 \hZ_{\bn,\tn}(\{\lambda^+\}|\{x\})
                 \tZ_{\bn,\tn}(\{\lambda^+\}|\{x\})}
                {\prod_{j=1}^{n}\mfa'(\lambda_j^+)
                 \prod_{j=1}^{\bm-\bn}(1+\mfa(\bx_j^-))
                 \prod_{j=1}^{\tm-\tn}(1+\mfab(\tx_j^-))}  \nn \\
&\qquad=\frac{1}{n!}\int_{(\mathcal{C}-\bGamma-\tGamma)^n}
 \prod_{j=1}^n\left[\frac{\d \omega_j}{2\pi\i(1+\mfa(\omega_j))} \right]
      \frac{Y_{\bn,\tn}(\{\omega\}|\{x^+\})
                 \hZ_{\bn,\tn}(\{\omega\}|\{x\})
                 \tZ_{\bn,\tn}(\{\omega\}|\{x\})}
                {\prod_{j=1}^{\bm-\bn}(1+\mfa(\bx_j^-))
                 \prod_{j=1}^{\tm-\tn}(1+\mfab(\tx_j^-))}. 
\label{mult-int1}
\end{align}
%
%
%
%

Next step is to reduce the integrals along the contour
$\mathcal{C}-\bGamma-\tGamma$
to those along the canonical contour $\mathcal{C}$.
We first consider the integrals on $\bGamma$. Because
the integrand in \eqref{mult-int1} is symmetric with
respect to $\{\omega\}$, we can divide 
the integral
\beq
\int_{(\mathcal{C}-\bGamma-\tGamma)^n}
\prod_{j=1}^n \frac{\d \omega_j}{2\pi\i}
\longrightarrow
\sum_{k=1}^{\bn}(-1)^k\binom{n}{k}
\int_{(\mathcal{C}-\tGamma)^{n-k}}
\prod_{j=1}^{n-k} \frac{\d\omega_j}{2\pi \i}
\int_{\bGamma^k}\prod_{j=1}^{k}  \frac{\d \omega_{n-k+j}}{2\pi \i},
\label{seperation}
\eeq
where $n=\bn+\tn$. Note that the sum over $k$ is
restricted to $k\le \bn$, since $|\bx|=\bn$ and the 
integrand vanishes when any two of $\{\omega\}$ 
is the same. Noting that, inside $\bGamma$, 
$G(\omega_j,\bx_k)$ \eqref{G-function} has simple poles 
$\omega_j=\bx_k$ with residues $-1$, and the poles
at $\omega_j=\bx_k$ for $\bM(\omega_j,\bx_k)$ \eqref{M-function}
are canceled by simple zeros of $\bmfb_+(\omega)$, we have
\begin{align}
&\int_{\bGamma^k} \prod_{j=1}^{k}\left[   
           \frac{\d \omega_{n-k+j}}{2\pi \i(1+\mfa(\omega_j))}
                \right]
    Y_{\bn,\tn}(\{\omega_j\}_{j=1}^n|\{x^+\})
                 \hZ_{\bn,\tn}(\{\omega_j\}_{j=1}^n|\{x\})
                 \tZ_{\bn,\tn}(\{\omega_j\}_{j=1}^n|\{x\})\nn \\
&=k! \sum_{\substack{\{\bx^{+-}\}\cup\{\bx^{++}\}=\{\bx^+\}\\
                     \{\bx^{++}\}\cup\{\tx^+\}=\{x^{++}\} \\
                     |\bx^{+-}|=k}}
          \frac{ Y_{\bn-k,\tn}(\{\omega_j\}_{j=1}^{n-k}|\{x^+\})}
               {\prod_{j=1}^k (1+\mfa(\bx^{+-}_j))}    
    \prod_{j=1}^{\tm-\tn}
\left[
  1+\kappa^{-1} \mfab(\tx_j^-)\prod_{l=1}^{n-k} 
                          \frac{f(x_l^{++},\tx_j^-)f(\tx_j^-,\omega_l)}
                               {f(\tx_j^-,x_l^{++})f(\omega_l,\tx_j^-)} 
  \right]   \nn \\
&
 \times \prod_{j=1}^{\bm-\bn}
     \left[
        1+\kappa\mfa(\bx_j^-)\prod_{l=1}^{n-k} 
                  \frac{f(\bx_j^-,x_l^{++})f(\omega_l,\bx_j^-)}
                       {f(x_l^{++},\bx_j^-)f(\bx_j^-,\omega_l)}                       
       \right] 
 \prod_{j=1}^{k}
     \left[
         1-\kappa 
            \prod_{l=1}^{n-k}\frac {f(\bx_j^{+-},x_l^{++})
                               f(\omega_l,\bx_j^{+-})}
                           {f(x_l^{++},\bx_j^{+-})
                                   f(\bx_j^{+-},\omega_l)}
     \right].
\end{align}
Performing a resummation as in \cite{GKS04},
one obtains
\begin{align}
\Phi_N(\kappa|\{\xi\})= &\sum_{\bn=0}^{\bm}\sum_{\tn=0}^{\tm}
                   \sum_{\substack{\{\bx^+\}\cup\{\bx^-\}=\{\bx\} \\                                                          |\bx^+|=\bn}}
                   \sum_{\substack{\{\tx^+\}\cup\{\tx^-\}=\{\tx\} \\                                                          |\tx^+|=\tn}}
      \frac{\kappa^{\bm-\tm-\bn}}{n!} \prod_{j=1}^n 
          \frac{1}{\bmfb'_-(x_j^+)\tmfb'_+(x_j^+)}\nn \\
 &\times \int_{(\mathcal{C}-\tGamma)^{n}}\prod_{j=1}^n
        \left[ 
          \frac{\d \omega_j \bmfb_-(\omega_j)\tmfb_+(\omega_j)}
              {2\pi\i(1+\mfa(\omega_j))}
          \prod_{k=1}^{\bn}\frac{\sh(\omega_j-\bx_k^+-\eta)}{\sh(x_j^+-\bx_k^+-\eta)}
                \prod_{k=1}^{\tn}\frac{\sh(\omega_j-\tx_k^++\eta)}
                                      {\sh(x_j^+-\tx_k^++\eta)}
        \right]   \nn \\
&
  \times\det M(\omega_j,x^+_k) \det G(\omega_j,x^+_k) \nn \\
&
  \times\prod_{k=1}^{\tm-\tn}\frac{1}{1+\mfa(\tx_k^-)}
   \prod_{k=1}^{\tm-\tn}\left[1+\kappa\mfa(\tx_k^-)
    \prod_{j=1}^n \frac{f(\omega_j,\tx_k^-)f(\tx_k^-,x_j^+)}
                       {f(\tx_k^-,\omega_j)f(x_j^+,\tx_k^-)}
   \right],
\end{align}
where
\beq
M(\omega_j, x_k^+)=t(x_k^+,\omega_j) 
            \prod_{l=1}^n\frac{\sh(\omega_j -x_l^+-\eta)}
                              {\sh(\omega_j -\omega_l-\eta)}
           +\kappa t(\omega_j,x_k^+)
            \prod_{l=1}^n\frac{\sh(\omega_j -x_l^++\eta)}
                              {\sh(\omega_j -\omega_l+\eta)}.
\eeq

Now we would like to consider the integrals
along the contour $\tGamma$. Utilizing the
relation similar to \eqref{seperation}, we can
separate the $\mathcal{C}$-integrals from 
the $\tGamma$-integrals. 
The unwanted terms including $\mfa(\tx_j)$ etc,
however, cannot be eliminated by actually performing the 
$\tGamma$-integrals, since the poles inside $\tGamma$
are not $\{\tx\}$ but  $\{\tx-\eta\}$. Nevertheless,
we observe that all the unwanted terms vanish in the 
homogeneous limit $\xi_j\to 0$, because the auxiliary 
function $\mfa(\lambda)$ has poles (respectively zeros) of 
order $N_0/2+N_1+1$ at $\lambda=0$ (respectively 
$\lambda=\eta$). 
{}From the relation $\lim_{\{\xi\}\to\{0\}}G(\omega_j,\tx_k)=-
\lim_{\{\xi\}\to\{0\}}G(\omega_j,\tx_k-\eta)$
which is derived by adapting $\mfa(0)=\infty$, $\mfa(\eta)=0$ 
and $t(\omega,x)=t(x-\eta,\omega)$ to \eqref{G-function}, we
obtain
\begin{align}
&\Phi_N(\kappa|\{0\})= \lim_{\{\xi\}\to\{0\}}
       \sum_{\bn=0}^{\bm}\sum_{\tn=0}^{\tm}
                   \sum_{\substack{\{\bx^+\}\cup\{\bx^-\}=\{\bx\} \\                                                          |\bx^+|=\bn}}
                   \sum_{\substack{\{\tx^+\}\cup\{\tx^-\}=\{\tx\} \\                                                          |\tx^+|=\tn}}
      \frac{\kappa^{m-n}}{n!} \prod_{j=1}^n 
          \frac{1}{\bmfb'_-(x_j^+)\tmfb'_+(x_j^+)}\nn \\
 &\quad \times \int_{\mathcal{C}^{n}}\prod_{j=1}^n
        \left[ 
          \frac{\d \omega_j \bmfb_-(\omega_j)\tmfb_+(\omega_j)}
               {2\pi\i(1+\mfa(\omega_j))}
           \prod_{k=1}^{n}\frac{\sh(\omega_j-x_k^+-\eta)}
                                 {\sh(x_j^+-x_k^+-\eta)}
        \right]\det M(\omega_j,x^+_k) \det R(\omega_j,x^+_k).
\label{Phi-final}
\end{align}
Here the function $R(\omega_j,x_k^+)$ is defined by
\beq
R(\omega_j,x_k^+)=\begin{cases}
            G(\omega_j,x_k^+) & \text{for $x_k^+\in \{\bx\}$} \\
            -\kappa  G(\omega_j,x_k^+-\eta) 
                 \prod_{l=1}^{n}\frac{\sh(x_k^+-\omega_l-\eta)\sh(x_k^+-x_l^++\eta)}
                                      {\sh(x_k^+-\omega_l+\eta)\sh(x_k^+-x_l^+-\eta)}
                        & \text{for $x_k^+\in \{\tx\}$}
            \end{cases},
\eeq
and we have used $\bm-\tm=m$ and $\bn+\tn=n$.
The integrand of \eqref{Phi-final} is a symmetric function
of  $\{x^+\}$ and vanishes at $\bx^+_j=\bx^+_k$ and 
$\tx^+_j=\tx^+_k$. Thanks to this together with
the fact that $1/\bmfb_{-}(\lambda)$ (respectively 
$1/\tmfb_+(\lambda)$) has simple poles at $\lambda=\bx$
(respectively $\lambda=\tx$), we can directly apply \eqref{sum2int}
to \eqref{Phi-final}. Thus, we  finally arrive at
\begin{align}
\Phi_N(\kappa|\{0\})= &\lim_{\{\xi\}\to\{0\}}
       \sum_{n=0}^{\bm+\tm}
      \frac{\kappa^{m-n}}{(n!)^2} \prod_{j=1}^n 
\left[
     \int_{\Gamma\{0 \}\cup\Gamma\{\eta\}}   
      \frac{\d\zeta_j}{2\pi\i \bmfb_-(\zeta_j) \tmfb_+(\zeta_j)}
     \int_{\mathcal{C}}        
          \frac{\d \omega_j \bmfb_-(\omega_j)\tmfb_+(\omega_j)}
               {2\pi\i(1+\mfa(\omega_j))}
\right] \nn \\
& \times
           \prod_{j,k=1}^{n}\frac{\sh(\omega_j-\zeta_k-\eta)}
                               {\sh(\zeta_j-\zeta_k-\eta)}
  \det M(\omega_j,\zeta_k) \det R(\omega_j,\zeta_k),
\label{Phi-final2}
\end{align}
where the contour $\Gamma\{0\}\cup\Gamma\{\eta\}$ 
surrounds the points $0$ and $\eta$ and does not
contain any other singularities. In \eqref{Phi-final2},
the Trotter limit $N_0\to\infty$, $N_1\to\infty$ and
the limit $\ep\to 0$ can be taken analytically.
In this limit,  the auxiliary function $\mfa(\lambda)$
is  the solution to the nonlinear integral equation
\eqref{nlie1}.
Substituting \eqref{inhom} and \eqref{Trotter}, we
easily obtain that
\begin{align}
\lim_{N_1\to\infty}\lim_{\substack{\xi\to0\\
                \ep\to 0}}
\bmfb_-(\lambda)\tmfb_+(\lambda)
&=\lim_{N_1\to\infty}\left[\frac{\sh(\lambda+\beta_1/N_1)}
                 {\sh(\lambda+\beta_1/N_1-\eta)} \right]^{\frac{N_1}{2}}
                        \left[\frac{\sh(\lambda-\eta)}{\sh(\lambda)} 
                        \right]^{\frac{N_1}{2}} 
                  \left[\frac{\sh\lambda}
                 {\sh(\lambda-\eta)} \right]^{m}\nn \\
&=\exp\left[-\i t e\(\lambda-\frac{\eta}{2}\)-\i m p\(\lambda-\frac{\eta}{2}\)\right],
\end{align}
where $e(\lambda)$ and $p(\lambda)$ are related to 
the bare one-particle energy and momentum,
respectively. They are explicitly  given by
\beq
e(\lambda)=\frac{2J\sh^2\eta}{\sh(\lambda+\frac{\eta}{2})
                             \sh(\lambda-\frac{\eta}{2})},
\quad
p(\lambda)=\i\ln\(\frac{\sh(\lambda+\frac{\eta}{2})}{\sh(\lambda-\frac{\eta}{2})}\).
\label{EP}
\eeq
{}From \eqref{Phi-final2}, \eqref{Phi} and \eqref{g-function},
we end up with the following theorem.
\begin{theorem}
The generating function of the longitudinal dynamical correlation function
has the following multiple integral representation:
\begin{align}
&\mathcal{Q}_\kappa(m,t)=
 \sum_{n=0}^{\infty}
      \frac{\kappa^{m-n}}{(n!)^2} \prod_{j=1}^n 
\left[
     \int_{\Gamma\{0\}\cup\Gamma\{\eta\}}   
      \frac{\d\zeta_j}{2\pi\i}
     \int_{\mathcal{C}}        
          \frac{\d \omega_j}
               {2\pi\i(1+\mfa(\omega_j))}
\right] \prod_{j,k=1}^{n}\frac{\sh(\omega_j-\zeta_k-\eta)}
                               {\sh(\zeta_j-\zeta_k-\eta)} \nn \\
& \quad \times
   \prod_{j=1}^n \e^{\i t(e(\zeta_j-\frac{\eta}{2})-e(\omega_j-\frac{\eta}{2}))
                     +\i m(p(\zeta_j-\frac{\eta}{2})-p(\omega_j-\frac{\eta}{2}))}
  \det M(\omega_j,\zeta_k) \det R(\omega_j,\zeta_k).
\label{mult}
\end{align}
In this expression the function $\mfa(\lambda)$ is 
the solution to the nonlinear integral equation \eqref{nlie1}; 
$e(\lambda)$ and $p(\lambda)$, respectively, denote
the bare one-particle energy and momentum \eqref{EP}; the
function $M(\omega,\zeta)$ and $R(\omega,\zeta)$
are, respectively, defined by

\begin{align}
M(\omega, \zeta)&=t(\zeta,\omega) 
            \prod_{l=1}^n\frac{\sh(\omega -\zeta_l-\eta)}
                              {\sh(\omega -\omega_l-\eta)}
           +\kappa t(\omega,\zeta)
            \prod_{l=1}^n\frac{\sh(\omega -\zeta_l+\eta)}
                              {\sh(\omega -\omega_l+\eta)}, \nn \\
R(\omega,\zeta)&=
\begin{cases}
            G(\omega,\zeta) & \text{for $\zeta\sim 0$} \\
            -\kappa G(\omega,\zeta-\eta) 
                 \prod_{l=1}^{n}\frac{\sh(\zeta-\omega_l-\eta)\sh(\zeta-\zeta_l+\eta)}
                                      {\sh(\zeta-\omega_l+\eta)\sh(\zeta-\zeta_l-\eta)}
                        & \text{for $\zeta\sim \eta$}
            \end{cases},
\end{align}
where $t(\lambda,\mu)=\sh(\eta)/(\sh(\lambda-\mu)\sh(\lambda-\mu+\eta))$ and
$G(\lambda,\zeta)$ is the solution of the linear integral equation
\beq
G(\lambda,\zeta)=t(\zeta,\lambda)+\int_{\mathcal{C}} \frac{\d \omega}{2\pi \i}
        \frac{\sh(2\eta)}{\sh(\lambda-\omega+\eta)\sh(\lambda-\omega-\eta)}
  \frac{G(\omega,\zeta)}{1+\mfa(\omega)}.
\label{G-function2}
\eeq
The contour $\mathcal{C}$ is the canonical contour (see figure~\ref{contour-pic})
and $\Gamma\{0\}\cup\Gamma\{\eta\}$ encircles the points
$0$ and $\eta$.
\end{theorem}

Thus the time-dependent longitudinal correlation function
$\bra \sigma_1^z(0)\sigma_{m+1}^z(t) \ket$ can be evaluated
by inserting the generating function \eqref{mult} into 
\eqref{op-Q2}, and using the magnetization
\begin{equation}
\bra \sigma^z \ket=-1-\int_{\mathcal{C}}\frac{\d\omega}{\pi \i}
                           \frac{G(\omega,0)}{1+\mfa(\omega)}
                  =1+\int_{\mathcal{C}}\frac{\d\omega}{\pi \i}
                           \frac{G(\omega,0)}{1+\mfab(\omega)},
\label{magnet2}
\end{equation}
which is derived from \eqref{nlie1}, \eqref{nlie2},
\eqref{magnet} and \eqref{G-function2}
\cite{GKS04}.
%
%
%
\section{Special cases}
%
Here we  comment on some special cases 
derived from the multiple integral representation \eqref{mult}
for the dynamical correlation function  
$\bra \sigma_1^z(0)\sigma_{m+1}^z(t)\ket$.

\subsection{Static limit $t=0$}

First let us consider the static limit $t=0$. Due to
the factors $e(\zeta_j-\eta/2)$, the integrand in \eqref{mult}
has essential singularities located on $\zeta_j=0$ and $\eta$.
In the static limit these essential 
singularities vanish, and hence the integrals along
the contour $\Gamma\{\eta\}$ disappear. Because
the remaining part of the integrand has poles of order
$m$ at the points $\zeta_j=0$, the integrals on the 
contour $\Gamma\{0\}$ vanish for $n> m$.
Namely the sum over $n$ is restricted to $n\le m$.
The resultant expression reads
\begin{align}
\mathcal{Q}_\kappa(m,0)=&
 \sum_{n=0}^{m}
      \frac{\kappa^{m-n}}{(n!)^2} \prod_{j=1}^n 
 \left[
     \int_{\Gamma\{0\}}   
          \frac{\d\zeta_j}{2\pi\i}
             \left\{\frac{\sh(\zeta_j-\eta)}{\sh(\zeta_j)}\right\}^m
     \int_{\mathcal{C}}        
          \frac{\d \omega_j}
               {2\pi\i(1+\mfa(\omega_j))}
             \left\{\frac{\sh(\omega_j)}{\sh(\omega_j-\eta)}\right\}^m
  \right] \nn \\
 &\times \prod_{j,k=1}^{n}\frac{\sh(\omega_j-\zeta_k-\eta)}
                               {\sh(\zeta_j-\zeta_k-\eta)} 
  \det M(\omega_j,\zeta_k) \det G(\omega_j,\zeta_k),
\end{align}
agreeing with that in \cite{GKS04}.

\subsection{ Zero-temperature limit $T=0$}

For direct comparison with the previous result for 
$T=0$ \cite{KMST05}, it is
convenient to introduce another expression of the
generating function.  Considering the correlation function
$\bra \sigma_{m+1}^z(-t) \sigma_1(0)\ket$ which is
equivalent to $\bra\sigma^z_1(0)\sigma^z_{m+1}(t) \ket$,
and utilizing the technique described in the previous
section\footnote{In this case, one needs to modify
the definition of the spectral parameters in the
quantum transfer matrix as $\lambda \pm \varepsilon_j \to \lambda$
and $-\lambda\to -\lambda\pm \varepsilon_j$, where
$\varepsilon_j=\varepsilon, \varepsilon_0, \varepsilon_1$.}, one may obtain
\begin{align}
&\mathcal{Q}_\kappa(m,t)=
 \sum_{n=0}^{\infty}
      \frac{(-1)^n}{(n!)^2} \prod_{j=1}^n 
\left[
     \int_{\Gamma\{-\eta\}\cup\Gamma\{0\}}   
      \frac{\d\zeta_j}{2\pi\i}
     \int_{\mathcal{C}}        
          \frac{\d \omega_j}
               {2\pi\i(1+\mfab(\omega_j))}
\right] \prod_{j,k=1}^{n}\frac{\sh(\omega_j-\zeta_k+\eta)}
                               {\sh(\zeta_j-\zeta_k+\eta)} \nn \\
& \quad \times
   \prod_{j=1}^n \e^{\i t(e(\zeta_j+\frac{\eta}{2})-e(\omega_j+\frac{\eta}{2}))
                     +\i m(\bp(\zeta_j+\frac{\eta}{2})-\bp(\omega_j+\frac{\eta}{2}))}
  \det M(\omega_j,\zeta_k) \det \overline{R}(\omega_j,\zeta_k),
\label{mult2}
\end{align}
where $\bp(\lambda)$ and $\overline{R}(\omega,\zeta)$  are defined by
\begin{align}
&\bp(\lambda)=\i \ln \(\frac{\sh(\lambda-\frac{\eta}{2})}
                           {\sh(\lambda+\frac{\eta}{2})}\), \nn \\
&\overline{R}(\omega,\zeta)=
\begin{cases}
            G(\omega,\zeta)  &\text{for $\zeta\sim 0$} \\
            -\kappa^{-1}G(\omega,\zeta+\eta) 
                 \prod_{l=1}^{n}\frac{\sh(\zeta-\omega_l+\eta)\sh(\zeta-\zeta_l-\eta)}
                                      {\sh(\zeta-\omega_l-\eta)\sh(\zeta-\zeta_l+\eta)}
                        &\text{for $\zeta\sim-\eta$}
\end{cases}.
\end{align}
Note that $G(\lambda,\zeta)$ is the solution of the integral
equation \eqref{G-function} which is also written in terms of
$\mfab(\lambda)$ if $\lambda$ and $\zeta$ are located inside $\mathcal{C}$:
\beq
G(\lambda,\zeta)=-t(\lambda,\zeta)-\int_{\mathcal{C}} \frac{\d \omega}{2\pi \i}
        \frac{\sh(2\eta)}{\sh(\lambda-\omega+\eta)\sh(\lambda-\omega-\eta)}
  \frac{G(\omega,\zeta)}{1+\mfab(\omega)}.   
\label{G-function3}
\eeq
Here we restrict ourselves on  the off-critical case 
$\Delta>0$ and set $\eta<0$ as in \cite{KMST05}. 
Note that  we can also treat the critical 
case $|\Delta|\le 1$ by just changing the definition 
of the integration contour as in figure~\ref{contour-pic}.

Shifting the variables in \eqref{mult2} by  $\omega_j \to \omega_j-\eta/2$
and $\zeta_j\to\zeta_j-\eta/2$, we consider the integrals on the
contour $\Gamma\{-\eta/2\}\cup\Gamma\{\eta/2\}$ and $-\mathcal{C}_0\cup
\mathcal{C}_{\frac{\eta}{2}}$. Here $\mathcal{C}_0$ and $\mathcal{C}_{\frac{\eta}{2}}$
denote  $\mathcal{C}_0=[-\pi \i/2,\pi\i/2]$ and
$\mathcal{C}_{\frac{\eta}{2}}=[\eta/2-\pi \i/2,\eta/2+\pi \i/2]$, respectively.
By close analysis of the auxiliary function $\mfab(\lambda)$ for 
$h>0$ and $\eta <0$ at the
zero-temperature limit $T\to 0$, one finds 
\beq
\frac{1}{1+\mfab(\lambda)}\xrightarrow{T\to 0}\begin{cases}
                           0 &\text{for $\lambda\in\mathcal{C}_0 
                                         \cup \mathcal{C}_{\frac{\eta}{2}}\setminus
                                         \mathcal{L}$} \\
                           1 &\text{for $\lambda\in\mathcal{L}$}
                          \end{cases},
\quad \mathcal{L}\in \left[-q_h+\frac{\eta}{2},q_h+\frac{\eta}{2}\right],
\label{Tzero}
\eeq
where the `Fermi point' $q_h$ is an imaginary number depending on $h$.
Substituting this into \eqref{G-function3} and appropriately shifting
the variables, we have
\begin{equation}
G\(\lambda-\frac{\eta}{2},\zeta-\frac{\eta}{2}\)=
-t(\lambda,\zeta)+\int_{-\mathcal{L}} \frac{\d \omega}{2\pi \i}
        \frac{\sh(2\eta)G\(\omega-\frac{\eta}{2},\zeta-\frac{\eta}{2}\)}
              {\sh(\lambda-\omega+\eta)\sh(\lambda-\omega-\eta)}.
\end{equation}
Comparing above with equation (2.16) in \cite{KMST02}, one identifies 
$G(\lambda,\zeta)$ as the density function $\rho(\lambda,\zeta)$:
\begin{equation}
G\(\lambda-\frac{\eta}{2},\zeta-\frac{\eta}{2}\)
=2\pi \i \rho(\lambda,\zeta).
\label{density}
\end{equation}
Inserting \eqref{density} and \eqref{Tzero} into \eqref{mult2}, we arrive at
the expression for the generating function at $T=0$:
\begin{align}
\lim_{T\to0}\mathcal{Q}_\kappa(m,t)=&
 \sum_{n=0}^{\infty}
      \frac{1}{(n!)^2} \prod_{j=1}^n 
\left[
     \int_{\Gamma\{\pm \eta/2\}}   
      \frac{\d\zeta_j}{2\pi\i}
     \int_{-\mathcal{L}}        
        \d \omega_j
\right] \prod_{j,k=1}^{n}\frac{\sh(\omega_j-\zeta_k+\eta)}
                               {\sh(\zeta_j-\zeta_k+\eta)} \nn \\
& \quad \times
   \prod_{j=1}^n \e^{\i t(e(\zeta_j)-e(\omega_j))
                     +\i m(\bp(\zeta_j)-\bp(\omega_j))}
  \det M(\omega_j,\zeta_k) \det \mathcal{R}(\omega_j,\zeta_k)
\label{multzero}
\end{align}
with
\begin{equation}
\mathcal{R}(\omega,\zeta)=
\begin{cases}
            \rho(\omega,\zeta)  &\text{for $\zeta\sim 0$} \\
            -\kappa^{-1}\rho(\omega,\zeta+\eta) 
                 \prod_{j=1}^{n}\frac{\sh(\zeta-\omega_l+\eta)\sh(\zeta-\zeta_l-\eta)}
                                      {\sh(\zeta-\omega_l-\eta)\sh(\zeta-\zeta_l+\eta)}
                        &\text{for $\zeta\sim-\eta$}
\end{cases}.
\end{equation}
The above expression reproduces equation (6.17) in \cite{KMST05}.

\subsection{XY (free fermion) limit $\Delta=0$}

Along the method described in 
\cite{KMST05} (see also \cite{KMST-free,GS06} for
the static case), we would like
to study the XY limit, where 
$\bra \sigma_1^z(0)\sigma_{m+1}^z(t)\ket$ can be written as
a product of single integrals. Set $\eta=\pi \i/2$. Then
the kernels of the integral equation in $\eqref{nlie1}$
and in $\eqref{G-function2}$ are equal to zero. Hence
\beq
\mfa(\lambda)=\exp\left[-\frac{h}{T}-
      \frac{1}{T}\frac{4 \i J}{\sh(2\lambda)}\right],
      \quad G(\lambda,\zeta)=\frac{-2}{\sh(2(\lambda-\zeta))}.
\eeq
Shifting the variables $\zeta_j\to\zeta_j+\pi \i/4$ and
$\omega_j\to\omega_j+\pi \i/4$ in \eqref{mult}, one easily
sees
\begin{align}
\mathcal{Q}_{\kappa}(m,t)=
\sum_{n=0}^{\infty} &
      \frac{\kappa^{m-n}}{(n!)^2} \prod_{j=1}^n 
\left[
     \int_{\Gamma\{\pm\pi \i/4\}}   
      \frac{\d\zeta_j}{2\pi\i}
     \int_{\mathcal{C'}}        
          \frac{\d \omega_j}
               {2\pi\i(1+\mfa(\omega_j+\frac{\pi}{4}\i))}
\right] \prod_{j,k=1}^{n}\frac{\ch(\omega_j-\zeta_k)}
                               {\ch(\zeta_j-\zeta_k)} \nn \\
&\times
   \prod_{j=1}^n \e^{\i t(e(\zeta_j)-e(\omega_j))
                     +\i m(p(\zeta_j)-p(\omega_j))}
\det M(\omega_j,\zeta_k) \det R(\omega_j,\zeta_k),
\label{multfree}
\end{align}
where
$\mathcal{C'}=-\mathcal{C}_0\cup\mathcal{C}_{-\pi\i/2}$;
$\mathcal{C}_0=[-\infty,\infty]$; $\mathcal{C}_{-\pi\i/2}
=[-\pi\i/2-\infty,-\pi\i/2+\infty]$. In this case 
the function $R(\omega,\zeta)$ and $M(\omega,\zeta)$ are reduced
 to
\beq
R(\omega,\zeta)=\begin{cases}
         -\frac{2}{\sh(2(\omega-\zeta))} &\text{ for $\zeta\sim \pi \i/4$} \\
         -\kappa \frac{2}{\sh(2(\omega-\zeta))} &\text{ for $\zeta\sim -\pi \i/4$}
                 \end{cases},
\,\,
M(\omega,\zeta)=\frac{2(\kappa-1)}{\sh(2(\omega-\zeta))}
               \prod_{l=1}^n\frac{\ch(\omega-\zeta_l)}
                               {\ch(\omega-\omega_l)}.
\label{RM}
\eeq
In the above expression, we notice that $M(\omega,\zeta)$ is factorized 
and has the factor $\kappa-1$, which significantly 
simplifies the integral representation. After taking 
the second derivative with respect to $\kappa$ and 
setting $\kappa=1$, one observes all the terms 
$n>2$ vanish. Firstly we consider the case $n=2$.
Substituting \eqref{RM} into \eqref{multfree}, we
extract the term corresponding to $n=2$ 
(written as $\mathcal{Q}^{(2)}_{\kappa}(m,t)$). After 
differentiating with respect to $\kappa$ and setting
$\kappa=1$, one has
\begin{align}
&\partial^2_{\kappa} \mathcal{Q}^{(2)}_{\kappa}(m,t)\Bigr|_{\kappa=1} \nn \\
&\,\,=\frac{1}{32\pi^4}\prod_{j=1}^2 \left[
               \int_{\mathcal{C}'}\frac{\d\omega_j}{1+\mfa(\omega_j+\frac{\pi\i}{4})} 
               \int_{\Gamma\{\pm \pi \i/4\}}\d\zeta_j
                                \right]
        \det\left[\frac{\e^{\i t e(\zeta_k)+\i m p(\zeta_k)}}
                   {\sh(\omega_j-\zeta_k)}\right]
         \det\left[\frac{\e^{-\i t e(\omega_j)-\i m p(\omega_j)}}
                   {\sh(\omega_j-\zeta_k)}\right].
\end{align}
The integral on $\Gamma\{\pm\pi\i/4\}$ can be evaluated
by considering the residues outside the contour
$\Gamma\{\pm\pi\i/4\}$ i.e. at the points $\zeta_j=\omega_k$.
This leads to
\begin{equation}
\partial^2_{\kappa} \mathcal{Q}^{(2)}_{\kappa}(m,t)\Bigr|_{\kappa=1} 
=\frac{-1}{4\pi^2}\prod_{j=1}^2 \left[
               \int_{\mathcal{C}'}\frac{\d\omega_j}{1+\mfa(\omega_j+\frac{\pi\i}{4})} 
                                \right]
        \det\left[\frac{1-\e^{\i t(e(\omega_j)-e(\omega_k))+
                               \i m (p(\omega_j)-p(\omega_k))}}
                   {\sh(\omega_j-\omega_k)}\right].
\end{equation}
Changing the variables $\cosh(2\omega)=-1/\cos p$ and identifying
\begin{equation}
\d\omega=\frac{\ch(2\omega)}{2}\d p, \quad 
e(\omega)=4J \cos p, \quad 
\frac{\sqrt{\ch(2\omega_j)\ch(2\omega_k)}}{\sh(\omega_j-\omega_k)}
=\frac{1}{\sin\frac{1}{2}(p_j-p_k)},
\label{change}
\end{equation}
one obtains
\begin{equation}
\partial^2_{\kappa} \mathcal{Q}^{(2)}_{\kappa}(m,t)\Bigr|_{\kappa=1} 
=\frac{-1}{16\pi^2}\prod_{j=1}^2 \left[
               \int_{-\pi}^{\pi}\d p_j \vartheta(p_j)
                                \right]
        \det\left[\frac{1-\e^{4\i t J(\cos p_j-\cos p_k)+
                               \i m (p_j-p_k)}}
                   {\sin\frac{1}{2}(p_j-p_k)}\right],
\end{equation}
where $\vartheta(p)$ is the Fermi distribution function
\begin{equation}
\vartheta(p)=\frac{1}{1+\exp[-\frac{h}{T}+\frac{4J\cos p}{T}]}.
\end{equation}
Taking the lattice derivative over $m$, we obtain the following relation,
without explicit evaluation of the multiple integral:
\beq
2D_m^2 \partial^2_{\kappa} \mathcal{Q}^{(2)}_{\kappa}(m,t)\Bigr|_{\kappa=1} 
=\frac{1}{\pi^2}\left[ \int_{-\pi}^{\pi}\d p \vartheta(p) \right]^2-
\frac{1}{\pi^2}\left|\int_{-\pi}^{\pi}\d p \vartheta(p) \e^{4 \i t J\cos p+\i m p}
\right|^2.
\label{q2}
\eeq

Next let us compute the term
$\mathcal{Q}^{(1)}_{\kappa}(m,t)$ corresponding to
 $n=1$. Its explicit form is
\begin{align}
\mathcal{Q}^{(1)}_{\kappa}(m,t)=\frac{\kappa^{m-1}(\kappa-1)}{4\pi^2}&
\left[\int_{\Gamma\{-\pi\i/4\}}\d \zeta +\kappa\int_{\Gamma\{\pi\i/4\}}\d \zeta
\right]   \nn \\
& \times
 \int_{\mathcal{C}'}\frac{\d \omega}{1+\mfa(\omega+\frac{\pi\i}{4})} 
\frac{\e^{\i t (e(\zeta)-e(\omega))+\i m(p(\zeta)-p(\omega))}}{\sh^2(\omega-\zeta)}.
\end{align}
Evaluating the integral on $\Gamma\{\pm \pi \i/4\}$, one has
\begin{equation}
\left[\int_{\Gamma\{-\pi\i/4\}}\d \zeta +\int_{\Gamma\{\pi\i/4\}}\d \zeta
\right]   
\frac{\e^{\i t (e(\zeta)-e(\omega))+\i m(p(\zeta)-p(\omega))}}{\sh^2(\omega-\zeta)}
=2\pi [t e'(\omega)+mp'(\omega)].
\end{equation}
It immediately follows that
\begin{align}
2D_m^2\partial_{\kappa}^2\mathcal{Q}^{(1)}_{\kappa}(m,t)\Bigr|_{\kappa=1} =&
\frac{4}{\pi^2}
\int_{\Gamma\{-\pi\i/4\}}\d \zeta 
 \int_{\mathcal{C}'}\frac{\d \omega}{1+\mfa(\omega+\frac{\pi\i}{4})} 
\frac{\e^{\i t (e(\zeta)-e(\omega))+\i m(p(\zeta)-p(\omega))}}
                            {\ch(2\omega)\ch(2\zeta)} 
\nn \\
&\quad +\frac{4}{\pi}\int_{\mathcal{C}'}\frac{\d \omega}
                {1+\mfa(\omega+\frac{\pi\i}{4})}   p'(\omega).
\end{align}
The contour $\Gamma\{-\pi \i/4\}$ in the first term 
can be replaced by $\mathcal{C}'$. After changing
variables as in \eqref{change}, one arrives at
\begin{equation}
2D_m^2\partial_{\kappa}^2\mathcal{Q}^{(1)}_{\kappa}(m,t)\Bigr|_{\kappa=1} =
 \frac{1}{\pi^2}\int_{-\pi}^{\pi} \d p\vartheta(p)
\e^{-4\i t J \cos p- \i m p}
  \int_{-\pi}^{\pi} \d q \e^{4\i t J \cos q+ \i m q}
  -\frac{4}{\pi}\int_{-\pi}^{\pi}\d p\vartheta(p).
\label{q1}
\end{equation}
Sum up \eqref{q2}, \eqref{q1} and
 $2D_m^2\partial_{\kappa} \mathcal{Q}^{(0)}(m,t)|_{\kappa=1}=4$
which is trivially obtained from $ \mathcal{Q}^{(0)}(m,t)=\kappa^m$.
Then inserting the result into \eqref{op-Q2} and
using the magnetization (see \eqref{magnet2}):
\begin{equation}
\bra \sigma^z \ket=-1+\frac{1}{\pi}\int_{-\pi}^{\pi}
          \d p\vartheta(p),
\end{equation}
we finally obtain 
\begin{align}
\bra\sigma_1^z(0)\sigma_{m+1}^z(t) \ket
   =&\bra \sigma^z\ket^2-
\frac{1}{\pi^2}\left|\int_{-\pi}^{\pi}\d p \vartheta(p) 
\e^{4 \i t J\cos p+\i m p}\right|^2 \nn \\
&\quad +\frac{1}{\pi^2}\int_{-\pi}^{\pi} \d p\vartheta(p)
\e^{-4\i  t J \cos p- \i m p}
  \int_{-\pi}^{\pi} \d q \e^{4\i t J \cos q+ \i m q}.
\label{free-int}
\end{align}
The above expression coincides with the known result
as in \cite{CIKT93}\footnote{In fact the correlation function 
$\bra \sigma^z_{m+1}(t)\sigma^z_1(0)\ket$ is considered
in \cite{CIKT93}. The result is the same as
that of $\bra \sigma_1^z(0)\sigma_{m+1}^z(-t)\ket$.}.
Of course, \eqref{free-int} can also be derived by starting
from the
generating function defined in \eqref{mult2}.
%
\section{Conclusion}
%
In this paper, we have derived a multiple integral
representation for the time-dependent longitudinal
correlation function $\bra \sigma_1^z(0) \sigma_{m+1}^z(t) 
\ket$ at any finite temperature and finite magnetic field.
The formula reproduces the known results in the following
three limits:
(i) static limit, (ii) low-temperature limit 
and (iii) XY limit.

It will be very interesting to consider the following 
problems which still remain open. First, it is well-known 
that the long-distance asymptotics of correlation functions
at the low-energy region (namely $T=0$ or $T\ll J$)  
in the critical regime can be derived by a field 
theoretical argument (see \cite{KBIbook} for example). 
In contrast, our multiple integral representation
\eqref{mult} is valid for any finite temperature and 
interaction strength. The exact computations of 
the asymptotic behavior from \eqref{mult} beyond the  
field theoretical predictions are of importance.

Second, how do we evaluate and extract
the long-time asymptotics of the correlation 
function
$\bra \sigma_1^z(0)\sigma_{m+1}^z(t) \ket$ at 
{\it infinite} temperature? In this case the auxiliary 
function $\mfa(\lambda)$ in \eqref{mult} becomes quite 
simple: $\mfa(\lambda)=1$.
This problem is of interest in relation to  the 
issue of spin-diffusion in the spin-1/2 Heisenberg XXZ
chain (see, for example, \cite{FLS,FM,GKS06,Sirker06}). 

The third is how to apply our formula to the 
calculation of crucial physical quantities such
as the dynamical spin structure factor 
\cite{BKM02,Sato04,CHM05,CM05,Pereira06,CH06}, which can
be actually measured by a neutron scattering
experiment. In relation to this problem, finally we would like 
to comment on the  form factor expansion.
In fact, the multiple integral
representation for $T=0$ (see \eqref{multzero})
is directly connected to the form factor expansion \cite{KMST05,KMST05-1}.
On the other hand, for the finite temperature case,
it is not clear whether \eqref{mult} has a connection with
the form factor expansion, since
\eqref{mult} is  derived by the quantum transfer matrix 
acting not on the quantum space but on the auxiliary space.
The form factor expansion is an important tool to
investigate the dynamical properties of the system, 
and therefore  explicit expressions  at finite 
temperature are also  desired.
%
\section*{Acknowledgments}
%
The author would like to thank J. Sato and 
M. Shiroishi for 
fruitful discussions. He also acknowledges with 
thanks the hospitality of the theory group of 
the Australian National University, where part of 
this work was done.
This work is partially supported 
by Grants-in-Aid for Young Scientists (B) No.~17740248,
Scientific Research (B) No.~18340112 and 
(C) No.~18540341 from
the Ministry of Education, Culture, Sports, Science 
and Technology of Japan.

%

\end{document}